# Title: High-throughput validation of phase formability and simulation accuracy of Cantor alloys

Changjun Cheng[1,&], Daniel Persaud[1,&], Kangming Li[3], Michael J. Moorehead[4], Natalie Page[5], Christian Lavoie[6], Beatriz Diaz Moreno[7], Adrien Couet[8], Samuel E Lofland[5], Jason Hattrick-Simpers[1,2,9,10*]

1 Department of Materials Science and Engineering, University of Toronto, Toronto, ON, Canada

2 Acceleration Consortium, University of Toronto, Toronto, ON, Canada

3 Material Science and Engineering, King Abdullah University of Science and Technology, Thuwal, Kingdom of Saudi Arabia

4 Idaho National Laboratory, Idaho Falls, ID, United States

5 Department of Physics and Astronomy, Rowan University, Glassboro, NJ, United States

6 IBM T.J. Watson Research Center, Yorktown Heights, NY, United States

7 Canadian Light Source, Saskatoon, SK, Canada

8 University of Wisconsin-Madison, Madison, WI, United States

9 Schwartz Reisman Institute for Technology and Society, Toronto, ON, Canada

10 Vector Institute for Artificial Intelligence, Toronto, ON, Canada

& The authors contributed equally to this work

* Corresponding author: jason.hattrick.simpers@utoronto.ca

Abstract

High-throughput methods enable accelerated discovery of novel materials in complex systems such as high-entropy alloys, which exhibit intricate phase stability across vast compositional spaces. Computational approaches, including Density Functional Theory (DFT) and calculation of phase diagrams (CALPHAD), facilitate screening of phase formability as a function of composition and temperature. However, the integration of computational predictions with experimental validation remains challenging in high-throughput studies. In this work, we introduce a quantitative confidence metric to assess the agreement between predictions and experimental observations, providing a quantitative measure of the confidence of machine learning models trained on either DFT or CALPHAD input in accounting for experimental evidence. The experimental dataset was generated via high-throughput in-situ synchrotron X-ray diffraction on compositionally varied FeNiMnCr alloy libraries, heated from room temperature to ~1000 °C. Agreement between the observed and predicted phases was evaluated using either temperature-independent phase classification or a model that incorporates a temperature-dependent probability of phase formation. This integrated approach demonstrates where strong overall agreement between computation and experiment exists, while also identifying key discrepancies, particularly in FCC/BCC predictions at Mn-rich regions to inform future model refinement.

## Introduction

High-entropy alloys (HEAs) exhibit exceptional mechanical and high-temperature performance due to their characteristic high configurational entropy and stable solid-solution crystal structures through the “cocktail”, “sluggish-diffusion”, and lattice-distortion effects. [1,2]. For example, the FeNiMnCr system, a four-element derivative of the Cantor family, shows good phase stability, irradiation resistance, and mechanical ductility [3-5], as a promising candidate under extreme conditions. However, phase identification across broader compositional ranges is challenging [6]: multiple phases can form, and phase stability varies with synthesis techniques and processing temperatures. The vast synthesis parameter space comes with a complicated structure-property relationship. For example, the appearance of intermetallic or ordered phases (e.g., σ, B2, $L1_2$) significantly alters the microstructure and mechanical behavior [7]. While certain intermetallic-containing HEAs may show enhanced hardness or strength, these secondary phases may also deteriorate corrosion or oxidation resistance in aqueous or high-temperature environments [8,9]. The phases formed may vary significantly within narrow composition and temperature ranges. Thus, phase identification across the full compositional and temperature space remains a highly complex and critical task for HEA design.

Computational tools offer a promising route to explore phase formability and transformation behavior. Fundamental enthalpy and ordering insights can be determined from Density functional theory (DFT) [10,11], while calculation of phase diagrams (CALPHAD) provides equilibrium phase diagram predictions based on thermodynamic databases [12]. These computational methods can be further leveraged through the use of machine learning (ML) to build predictive models that interpolate trends found in a dense array of computationally calculated compo[13-15]. Despite good agreement between computational predictions and experiments in simpler systems [16], the fidelity of modeling complex HEAs remains uncertain, particularly since there are multiple variables and uncertain criteria in high-throughput experimental (HTE) validation of computational calculations. For instance, CALPHAD and DFT have predicted several phases in FeNiMnCr system though, some of those phases have yet to be detected in the samples fabricated by arc melting or additive manufacturing [17,18]. There has been limited discussion on the discrepancies in structure prediction, and the agreement and fidelity between a computational method and practical measurement have not been assessed.

HTE allows one to broadly assess the validity of these computational predictions [19,20]. However, most previous investigations have either mapped multiple compositions at a fixed temperature [10,11,21] or performed in-situ diffraction on a single composition [22,23], limited by the constraints of available equipment [24]. Recent developments in robotic systems integrated with synchrotron beamlines allow sequential in-situ X-ray diffraction (XRD) across composition libraries, permitting simultaneous exploration of both temperature and compositional space to identify various phase transformation tendencies [25]. Such efforts can yield sufficient results to provide meaningful assessment and recalibration of the degree of trust for specific computational predictions.

In this study, we employed *in-situ* synchrotron XRD to characterize phase evolution in FeNiMnCr HEA system across both compositional and temperature spaces. We compare those observations with predictions from CALPHAD- and DFT-based ML models, and we propose a metric that evaluates the agreement between computational predictions and experimental observations. This metric could also assess agreement across different datasets and weigh the likelihood of a theory being aligned with experiments, guiding future integrated frameworks in autonomous alloy discovery.

## Results and Discussions

During *in-situ* synchrotron X-ray diffraction (XRD), all samples exhibited three distinct stages: an as-deposited metastable structure, stabilized solid solution (SS), and silicide. The stabilized state included not only the crystallization from amorphous structure but also the transformation from intermetallic (IM) to solid solution (SS) phases. In the FeNiMnCr HEA system, the formed SS structure is observed as either/both face-centered cubic (FCC) or/and body-centered cubic (BCC) phase. According to the distinct phase transformation and categories of SS formation types (Figure 1a), the samples could be categorized into the types at FeNi-, Cr-, and Mn-rich regions, as well as FeNi/Cr- and Cr/Mn-transition regions. The amorphous FeNi-rich region (Figure 1b) crystallized into an FCC phase, and the transformation temperature increased when the FeNi concentration decreased. As the Cr concentration increased (Figure 1c), a BCC phase appeared during crystallization while the σ phase remained stable. Here, there was no obvious change in transformation temperature with Cr composition. Unlike the transformation due to crystallization in FeNi- and Cr-rich regions, the detected transition temperatures in Mn-rich regions (Figure 1d) indicate the onset of grain growth from a nanocrystalline (NC) state to BCC structure. At the interface between the Mn-rich and Cr-rich regions (Figures 1e, f), there are two distinct BCC phases. At even higher temperatures, the silicon diffusion from the substrate was considerable [23]. The formation of silicide phases was more pronounced than the reaction with oxygen, as there were no detectable oxides in the XRD patterns, and the variation in HEA composition resulted in differing silicide phases (Figures S4 and S5).

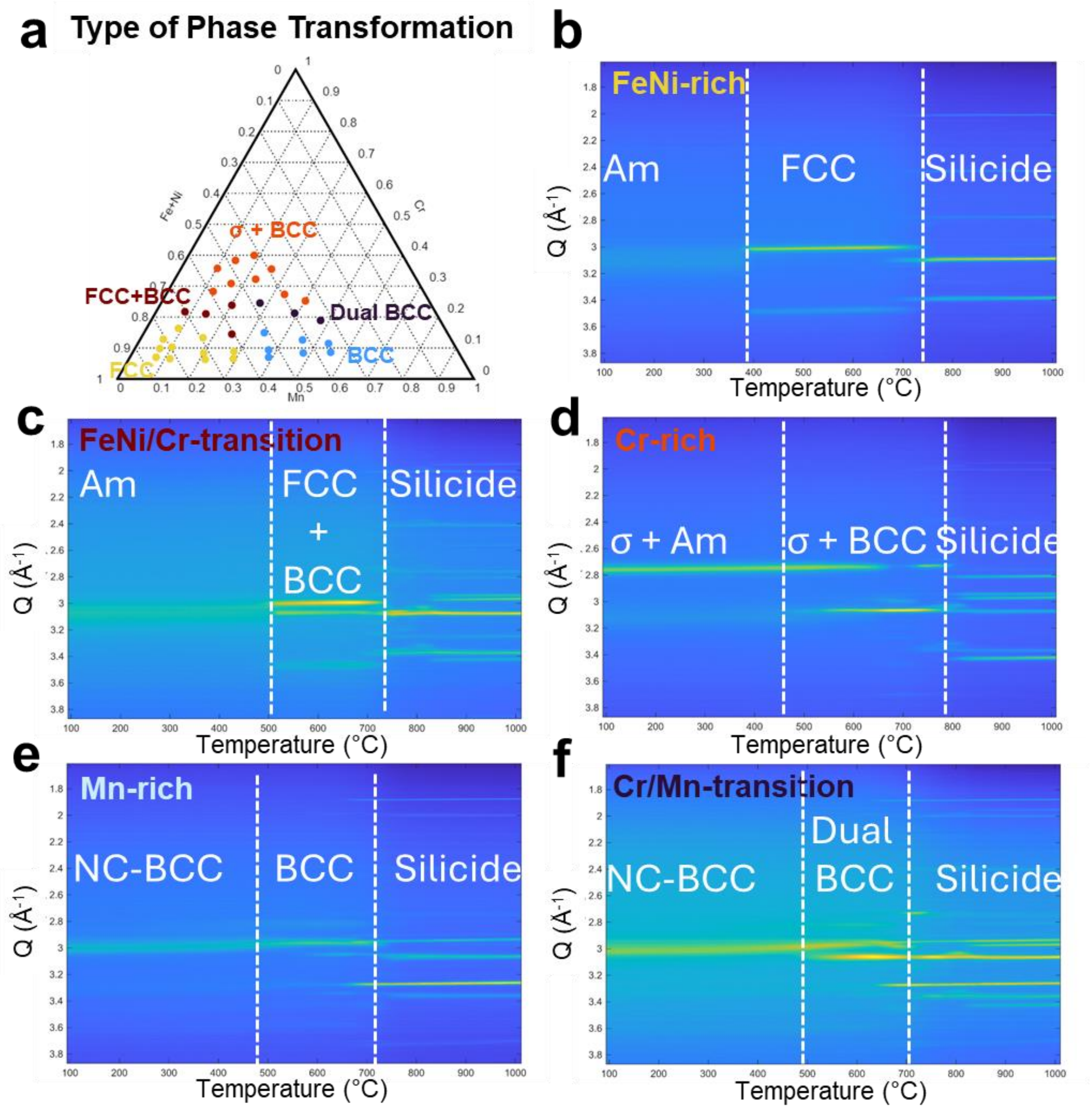

Figure 1. *In-situ* synchrotron X-ray diffraction results of all measured samples. (a) Types of phase transformation in pseudo pseudo-ternary diagram. (b-f) Corresponding XRD contour maps of phase transformation during heating.

Due to the relatively high deposition rate, all the samples exhibit NC or even amorphous structures in the as-deposited state [26-28]. Figure 2a shows the estimated grain size for the as-grown samples, determined from the Scherrer equation, applied to the peak width of the main reflection of each phase for a given composition. The as-deposited FeNi-rich samples showed the smallest grain size, corresponding to a fully amorphous structure. The as-deposited Mn-rich samples displayed an NC-BCC phase with a grain size above 3 nm. The as-deposited Cr-rich samples exhibited the coexistence of an amorphous phase with a σ phase, showing by far the largest crystallite size of any as-deposited samples.

As the samples were heated, all as-deposited structures remained unchanged below 300 °C, except the FeNi-richest sample with the transformation temperature $T_{tran}$ ~ 120 °C, and then transformed to thermodynamically stable phases at higher temperatures (Figure 2b). Cr-rich samples exhibit the highest crystallization temperature, and the transition temperatures in Mn-rich regions indicate the onset of grain growth from an NC state to stabilized BCC structures. The difference in phase formation is attributed to the variations in composition. In this HEA system, FeNi composition affects the formation tendency of BCC and FCC phases, and the σ phase only exists in the Cr-rich region. Similar to the prior study of FeNiCr alloys [29], the formation of the σ phase is attributed to a high Cr composition (Table S2). Figure 2c presents a clear formability boundary of the σ phase based on the FeNiMn/Cr ratio: the σ phase only exists when the value is lower than 3. As the FeNi concentration increases, i.e., with a larger FeNi/MnCr ratio, samples showed the tendency to form an FCC phase while the crystallization temperature decreased (Figure 2d). The samples with more Mn or Cr (FeNi/MnCr < 2.2) exhibited a preference to form a BCC phase, which could also transform into FCC phase at higher temperatures. Due to the overlap of FCC and silicide peaks (Figure S4), it is difficult to fully index the FCC phase, and obtaining an exact ratio range of FCC+BCC phases is challenging. Samples with a higher FeNi/MnCr ratio formed the FCC phase as the stable solid solution phase, however, with a lower crystallization temperature (inserted diagram in Figure 2d). A pure FCC phase existed with FeNi composition larger than 70%, i.e., FeNi/MnCr > 2.3. Overall, Fe and Ni favor FCC stabilization, whereas Cr and Mn drive BCC formation.

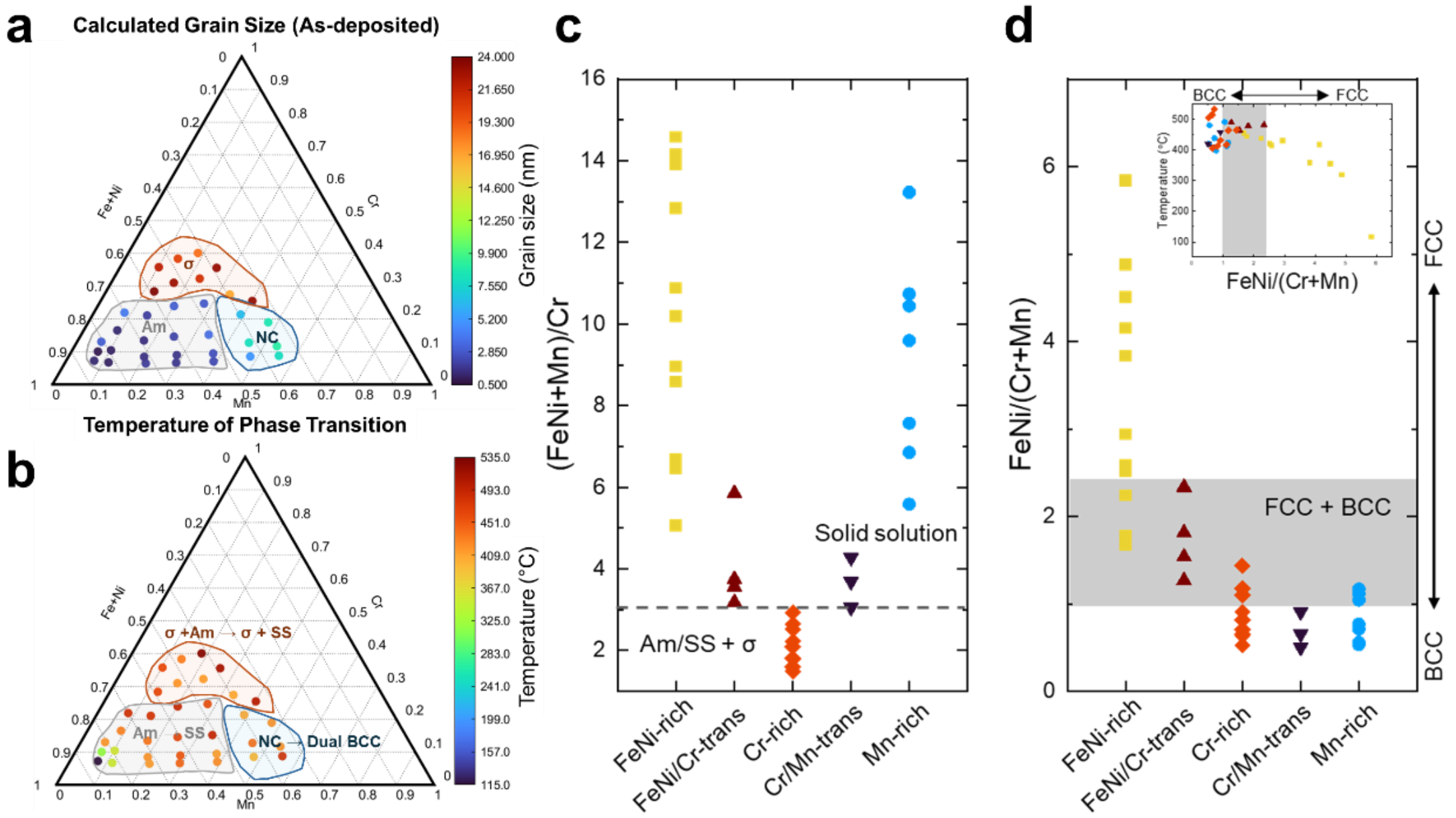

Figure 2. Summary of phase formability in the Cantor library. (a) Calculated grain size of the major phase for the as-deposited materials. (b) The onset temperature of phase transition and the major transformation reactions. The tendency of phase formation during heating based on (c)

FeNiMn/Cr and (d) FeNi/MnCr ratios. The inserted figure in (d) shows the phase transition temperature as a function of FeNi/MnCr ratio.

Besides the observed composition ratios from the experiment, we also evaluate the phase formability according to the empirical criteria [30] for HEAs (Figures 3a, 3b, and S6). The empirical omega-atomic radius mismatch ($\Omega$-$\delta$) relationship (Figure 3a) considers the equilibrium state of materials, and all FeNiMnCr samples were in the SS region, corresponding well to the phases in the stabilized state (500-700 °C). The existence of the σ phase in Cr-rich samples aligns with the criteria for intermetallic (IM) formation. It should be noted that the σ phase is observed in the characterization, and the IM phase indicates the predicted structure in empirical principles or computational simulations. Considering the as-deposited samples (inserted diagram in Figure 3a), the characterized amorphous phase existed in the region with a lower $\Omega$ value. It corresponds well to the criteria for forming metallic glass structures, though with a relatively higher $\Omega$value than that of the empirical rules. The electronegativity mismatch $\Delta\chi$ and the mixture valence electron concentration ($VEC_{mix}$) are common parameters to consider the formability of stable SSs (Figure 3b). The stable FCC phase exists when $VEC \geq 8.4$; the tendency to form the BCC phase increases when $VEC$ decreases [30,31]. Additionally, FCC tends to exist when $\Delta\chi$ is small while BCC forms with larger $\Delta\chi$ values, and a clear IM region exists in the medium $\Delta\chi$ range, where σ and Laves phases form with SS structures [32]. Hence, the phase stabilization trend aligns well with empirical rules.

The energy above the convex hull for FeNi-based quaternary HEAs over Fe-Ni-Mn-Cr-Co-Al-Si space was used as input in the Random Forest (RF) model to train a hybrid DFT-ML framework.[33,34] The total energy was computed at 0 K without considering vibrational entropy. The estimated transformation points from ordered intermetallic to disordered solid solution were calculated as a function of temperature from the ideal configurational entropy for the quaternary space (Figure S1). The result reveals a unique intermetallic formability tendency within the FeNiMnCr system compared to that of the other Cantor derivatives.

Figure 3c shows the predicted phase distribution in the FeNi-Mn-Cr ternary diagram using a DFT-RF model. The existence of the σ phase is apparent in Cr-rich region, and the predicted space for dual SS phases aligns with the observed FeNi/Cr-transition region. Most of the predicted SS regions match well with the measurements, except Mn-rich ones where the observed BCC region is predicted to be FCC. This might result from the relatively high effective cooling rate in magnetron sputtering, allowing for depositing thin films in a metastable state [26-28]. Similar to the amorphous structure in both FeNi- and Cr-rich regions, the NC-BCC structure is likely a metastable deposition product in the Mn-rich region. The driving force for BCC grain growth appears to be much lower than the BCC/FCC transformation barrier, leading to the existence of BCC phase during *in-situ* heating in Mn-rich region. This diagram provides a sequential phase classification (IM or SS, single or dual, FCC or BCC) according to the prediction after correction. The assumptions and threshold setting in the DFT-RF results generate a temperature-independent phase formation preference over the composition space.

The stability of SS phases is determined using DFT calculations by comparing their equilibrium energy, and broadly, these calculations agree with in-situ XRD results from 500-700 °C. However, due to the omission of vibrational entropy, it often overestimates the $T_{trans}$, compared to experimentally determined values, as suggested by prior work on Fe-Ni alloys [35,36]. Therefore, a correction factor is required for more accurate predictions of the $T_{trans}$, and the melting point $T_{MP}$ can be used as a normalization factor for samples with various compositions. Here, we propose the ratio of $T_{trans} / T_{MP}$ to better align the phase transformation tendency predicted by DFT to experimental observations for different compositions, where $T_{MP}$ is estimated by the rule of mixtures (Figure S8a). A lower factor value indicates a higher likelihood of forming either BCC or FCC SS phases. Using this factor, we can observe a rational threshold for whether a phase transition occurs. Based on the experimental observations, a value of 75% is determined to be suitable for the IM-SS classification, i.e., SS would form for samples with $T_{trans}$ lower than 75% $T_{MP}$. Similarly, a factor considering the transformation to either BCC or FCC phase is proposed by comparing $T_{trans}^{BCC}$ and $T_{trans}^{FCC}$. A threshold of 90% in the difference between $T_{trans}^{BCC}$ and $T_{trans}^{FCC}$ is observed for the coexistence of the dual SS phase where $T_{trans}^{BCC}$ and $T_{trans}^{FCC}$ represent the transition temperatures determined from the DFT-ML models for the BCC and FCC transitions, respectively.

CALPHAD simulations were performed for Fe-Ni-Mn-Cr quaternary systems at 400-700 °C (in 100°C increments) and 1000 °C [18,37]. The FCC, BCC, and σphase are the major structures observed over the composition space. As the temperature increases, the region of σ phase decreases, forming stable FCC or BCC SS structures. Figure S10 compares the CALPHAD-RF predicted equilibrium phases with the XRD-indexed phases in the 500-700 °C range. The best agreement between the CALPHAD-RF model (Figure 3d) and the experimental results occurs at 700 °C. Similar to the agreement in the Fe-rich region with the DFT-RF model, the CALPHAD-RF model also successfully predicts the existence of the FCC phase in the same composition region. However, the CALPHAD-predicted FCC area is larger than that of the experimental results, labeling the Mn-rich region as FCC and the Cr-rich region as dual SS phases, and the predicted FCC region still occupies a significantly large compositional space even at a higher temperature (1000 °C, Figure S10). In contrast to experimental observations, CALPHAD predicts that the σ phase should form in compositions with high Mn concentration. Therefore, despite excellent agreement in the FeNi-rich region, there are considerable discrepancies between CALPHAD predictions and experimental results in other regions of compositional space.

The distributions of the DFT-RF model predictions in the $\Omega$-$\delta$ plot and $\Delta\chi$-$VEC_{mix}$ diagrams are shown in Figures 3e and 3f, respectively. Considering the classification between IM and SS phases, DFT-RF results show a similar distribution to that of the empirical criteria in $\Omega$-$\delta$ plot, i.e., SS forms in the whole range, and IM is located in the upper region ($\sigma \cdot \Omega > 24$). We note that BCC prefers to exist in upper regions while FCC in lower ones in DFT-RF results, with a considerable area of overlapping transition. While the empirical relationship between $\Omega$ and $\delta$ does not differentiate the SS types, the $\Delta\chi$-$VEC_{mix}$ plot provides a more discriminating classification. Similar to experimental results, the predicted BCC tends to form when $VEC_{mix}$ is small, and the predicted IM is located in the medium $\Delta\chi$ range. However, a significant discrepancy exists in the low $VEC_{mix}$ and high $\Delta\chi$ region, i.e., Mn-rich and Cr/Mn-transition samples. These are the regions with the largest disagreement between computational and experimental results: the DFT-RF model predicts both types to present as FCC phase, but both experimental and empirical observations indicate a preference for BCC formation, while the CALPHAD-RF prediction labels Cr/Mn-transition samples as dual phase.

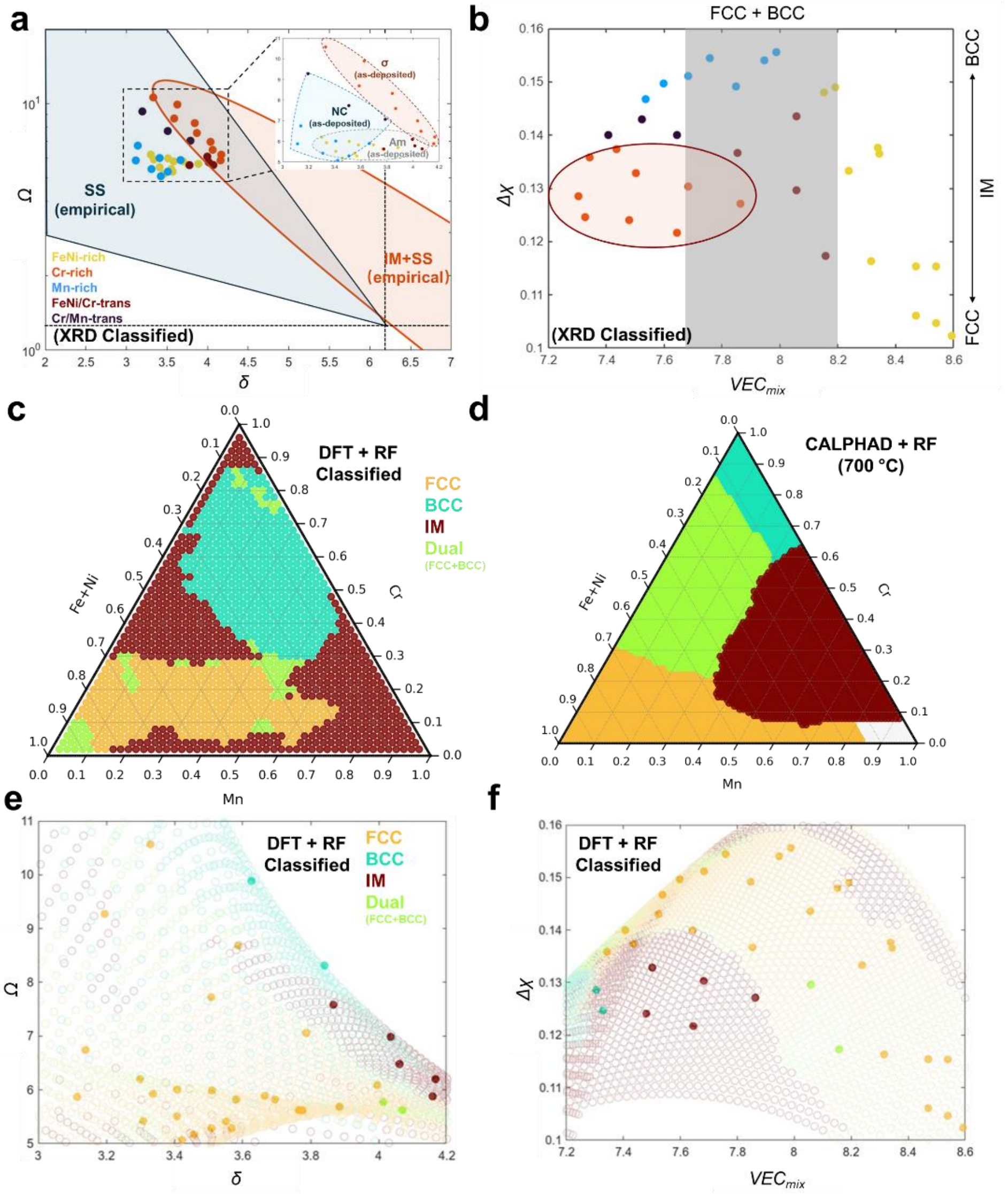

Figure 3. Phase formability based on empirical criteria and DFT calculation. (a) The distribution of different types of samples in the *Ω*-*δ* plot. The solid-line regions indicate the expected phases in empirical criteria, and the dashed-line regions in the inserted diagram show the phase types in as-deposited samples. (b) The distribution of samples in $\Delta\chi$-$VEC_{mix}$ plot. Ternary diagram of phase formation preference based on the hybrid (c) DFT-RF and (d) CALPHAD-RF models. The DFT-RF predicted results in (e) *Ω*-*δ* plot and (f) $\Delta\chi$-$VEC_{mix}$ plots, in which the solid and hollow points represent the sample and ternary space data, respectively.

As mentioned, the DFT-RF prediction provides a phase transformation tendency at each composition though, the temperature $T_{trans}$ for the transition determined from DFT calculations is not equivalent to the measured temperature of an actual sample. The above discussion describes a temperature-independent threshold to better align DTF-RF predictions to the experimental observations. However, this method is limited by correction based on observations, and the selection of threshold values only provides a temperature-independent phase classification. Rather than consider only the thermodynamically stable phase with temperature and composition, it may be more insightful to discuss $p_n(T)$ the formation probability of phase $n$. We model the phase transition following the classical logistic sigmoid function as it provides (1) clear class labels, (2) a convenient probability of the class, and (3) the ability to model abrupt phase transitions [38,39]. We propose to model the probability of the formation of the intermetallic phase $p_{IM}(T)$ with a modified logistic function:

$$p_{IM}(T) = \frac{1}{1 + e^{k(\Delta T^* - c)}}$$

where $\Delta T^* = (T - T_{trans}^{IM})/T_{MP}$ is the normalized difference in temperatures with $T_{trans}^{IM}$ being the predicted SS-IM transition temperature predicted by DFT, and $c$ a correction factor due to temperature overestimation. $k$ is a slope parameter, considering $\Delta T^*$values of all points

$$k = \frac{10}{max(\Delta T^*) - min(\Delta T^*)}.$$

Since the stable phases are either intermetallic or solid solution, the probability of SS formation $p_{SS}(T)$ is given by

$$p_{SS}(T) = 1 - p_{IM}(T)$$

The formation probability $p_{SS}(T)$of the SS phases encompasses both single and dual phase SSs (Figure S11), i.e.,

$$p_{SS}(T) = p_{SS}^S(T) + p_{SS}^D(T).$$

where $p_{SS}^S(T)$ and $p_{SS}^D(T)$ represent the probabilities of the single and dual phase SS, respectively. For the SS states, we assume that the relative formation of the dual SS state is temperature independent and model it as

$$p^* = \frac{1}{1+e^{\gamma}}$$

where $\gamma = k\left(\frac{|T_{trans}^{BCC} - T_{trans}^{FCC}|}{\min(T_{trans}^{BCC}, T_{trans}^{FCC})} - c^*\right)$. The threshold modification values of $c$ (0.25) and $c^*$ (0.1) are set based on the observation in Figure 3c. The temperature-dependent formation probabilities are:

$$p_{SS}^{D}(T) = p^*[1 - p_{IM}(T)]$$

$$p_{SS}^{S}(T) = (1 - p^*) \times p_{SS}(T) = (1 - p^*)[1 - p_{IM}(T)]$$

Thus, the probability of forming BCC or FCC (A or B) phase is:

$$p_A(T) = 1 - p_{IM}(T)$$

$$p_B(T) = p_{SS}^{D}(T)$$

for $T_{trans}^{A} < T_{trans}^{B}$ where $T_{trans}^{A}$ and $T_{trans}^{B}$ represent the predicted transition temperature of phase A and phase B, respectively.

Figure 4a shows the calibrated phase formability of the intermetallic structure at 700 °C. Besides Cr-based and Mn-based compositions (triangle corners), IM structure mainly exists at the Cr-rich region among the library, manifested as the σ phase in XRD. Figure 4b presents the temperature-independent $p_{SS}^{D}$ values in the composition space. The FeNi-rich, Mn-rich, and FeNi/Cr-transition regions also exhibited the tendency to form dual phases, while only FeNi/Cr-transition demonstrated the formation of FCC+BCC structure in XRD observations.

To assess how well models capture experimental behavior, we define the confidence metric $\kappa$, which considers the combination of both IM/SS and BCC/FCC pairs, i.e.,

$$\kappa = \frac{1}{2} * \sum_{i}^{\frac{SS}{IM}, \frac{BCC}{FCC}} A_i = \frac{1}{2} * (A_{IM} + A_{SS}) = \frac{1}{2} A_{IM} + \frac{1}{2}(0.5 * A_{BCC} + 0.5 * A_{FCC})$$

, where $A_i$ = 1 if the experimental observations for both phases of the pair agree with the predictions, 0.5 if only one phase agrees, and 0 if neither phase agrees. Thus, a higher value indicates a higher correlation of the predictions to the observations. Consider the sample in FeNi/Cr-transition region as an example (Figure 4c). The dual phases formed at 500 °C and were maintained above 700 °C. In the probability plot, the intersection of IM and BCC curves (~ 520 °C) indicates a considerable formation of the BCC phase; due to a high probability of dual phase, FCC is expected to form immediately after BCC. The good agreement between experiment and calculation results in a fully confident value ($\kappa$= 1) above 520 °C.

Since all samples present a relatively stable SS structure between 500 and 700 °C, we calculate the average $\kappa$values of DFT-RF models within this temperature range (Figure 4d). It should be noted that the co-existence of IM and SS was not considered when we set 0.5 as the phase formability threshold, which is slightly different from the true observations. The $\kappa$values of Cr-rich samples are lower than those of the FeNi-rich ones (~ 1). Despite the robust prediction of the non-existence of IM, the juxtaposition of the observed BCC and predicted FCC phases leads to a decrease in $\kappa$values for Mn-rich samples.

We also evaluate the CALPHAD-RF results to obtain the average $\kappa$ values over three temperatures (500, 600, and 700 °C) (Figure 4e). The FeNi-rich samples still exhibit the highest $\kappa$ value, but the experimentally observed BCC phase contradicts the predicted σ and FCC phases, resulting in the lowest $\kappa$values being located in the Mn-rich region. Therefore, despite CALPHAD-RF successfully predicting the structure in the FeNi-rich region, its agreement in other compositional regions is relatively low, with an especially considerable mismatch in the σ phase region compared to that of the DFT-RF model.

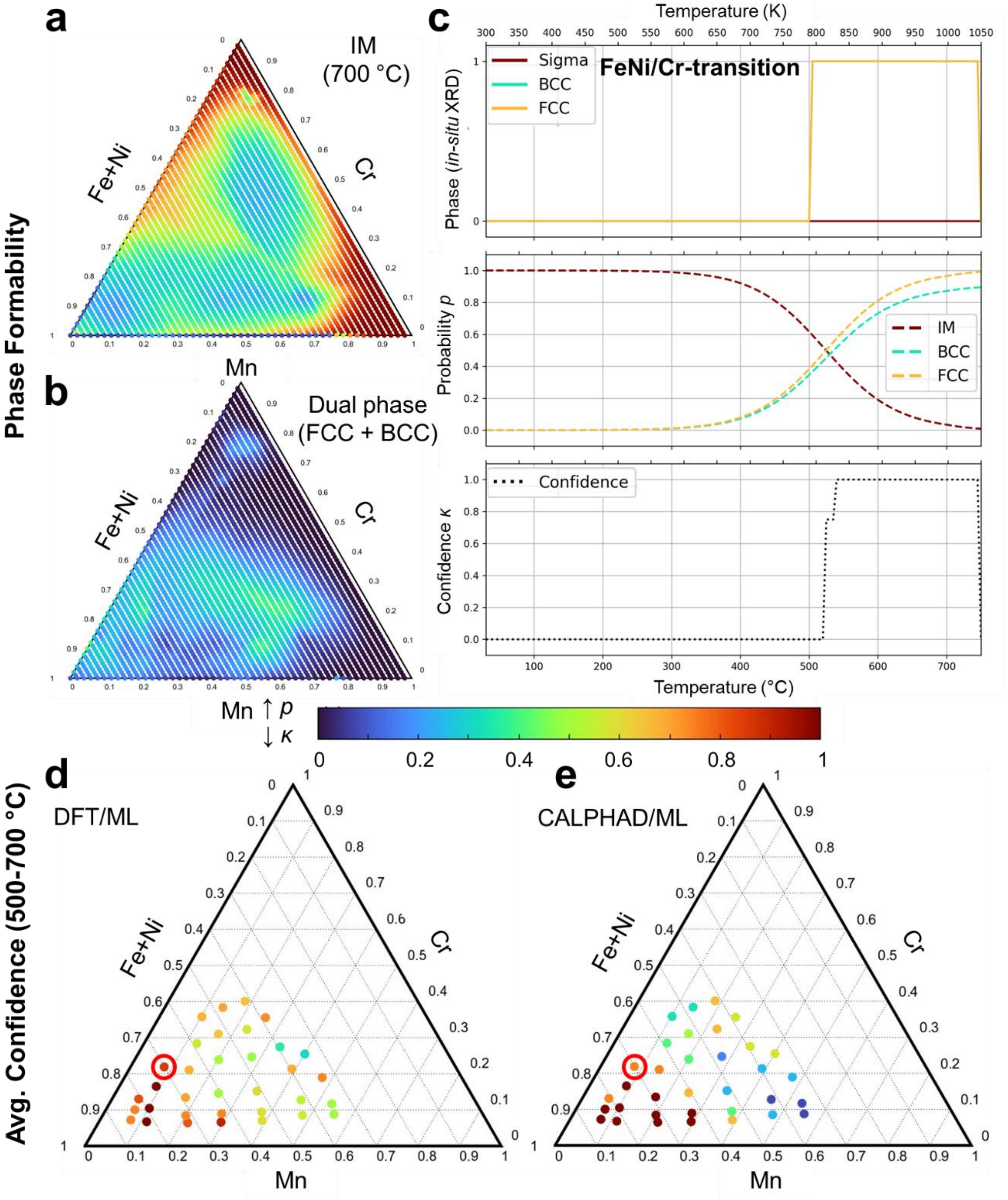

Figure 4. Phase formability and the agreement between calculation and experiment in FeNiMnCr HEAs. Phase formation probability of (a) IM at 700 °C and (b) dual phase structures based on the modified sigmoid calculation on DFT-RF results. (c) The comparison between XRD-indexed and DFT-RF predicted phases, as well as the corresponding confidence values as a function of temperature for the sample in FeNi/Cr-transition region. The average confidence values of (d) DFT-RF and (e) CALPHAD-RF models in the range of 500-700 °C, indicating the simulation

accuracy of the models. The point in the red circle indicates the sample in (c). The values in (a), (b), (d), and (e) correspond to the colors in the color bar.

The variability in values across the composition space highlights an opportunity to evaluate and refine the computational models in guiding experiments. Rather than relying solely on a fixed, sequential sampling approach, future iterations could integrate prior DFT and CALPHAD predictions with newly acquired experimental results to progressively update our assessment of model reliability. The automated experimental platform enables immediate incorporation of the outcomes of each iteration into an active learning framework with a modified acquisition function. This closed-loop strategy would prioritize experiments that yield maximum information, efficiently reducing uncertainty and effectively minimizing the required measurements. This approach has the potential to enhance the correspondence between simulations and experimental realities, improving the efficacy of high-throughput exploration in complex composition spaces.

## Conclusions

In this study, we systematically investigated the phase stability and transformation pathways of the FeNiMnCr HEA system by integrating high-throughput in-situ synchrotron XRD and combined computational DFT-RF and CALPHAD-RF methods. Our computational predictions revealed unique phase formability trends within the FeNiMnCr composition space, specifically highlighting the critical transition from IM to SS phases. Experimental validation via in-situ XRD across broad compositional and temperature ranges (room temperature to ~1000 °C) confirmed distinct phase transformations, including metastable-to-stable solid-solution transitions, as well as variation in FCC and BCC phases. To reconcile the discrepancies between predicted and observed transition temperatures, we introduced a modified logistic sigmoid function that calibrates predictions to experimental conditions and provides a probabilistic description of phase formation. We further proposed a confidence metric that quantifies the agreement between models and experiments across temperature and composition. After applying the modification to account for the temperature range of phase formation, DFT-RF predictions demonstrate better agreement with experimental results than the CALPHAD-RF model does, while discrepancies remained in the Mn-rich region.

## Method

DC magnetron co-sputtering was performed to fabricate FeNiMnCr HEA composition-gradient library (Figure S2a) with an AJA three-gun deposition system. High-purity 2-inch Cr and Mn targets were used in two of the cathodes. The third cathode contained overlapping semi-circular high-purity Fe and Ni foils. Deposition on 4-inch Si wafter was done with a pressure of 10 mTorr. The corresponding parameters are listed in Table S1. The *in-situ* synchrotron X-ray diffraction was performed in the Brockhouse beamline at Canadian Light Source (Figure S2b). Wafer samples were cut into small chips (12.5 mm × 16 mm) and arranged on Mo holders in the sample trays. KiNEDx Robot was applied to grab one of the samples and move it to the *in-situ* XRD instrument. The incidence X-ray was set as 7.16 keV with a wavelength of 1.7318 Å. The actual two-theta range of the collected XRD signal was 25.6° - 64.4°, and the equivalent angle covering the same *d* range in Cu Kα wavelength would be 22.7° - 56.6°. The average temperature ramp rate was 4 °C/s, and the corresponding rate variation of calibrated temperatures is shown in Figure S17. The XRD results experienced smoothing, background correcting, and peak fitting using Savitzky-Golay filter [40], asymmetric truncated quadratic [41], and Gaussian + Lorentzian model, respectively. Annealing was performed up to 1000 °C within purified He with a hot Ti charge. The temperature was determined from cubic spline interpolation of the known melting temperatures of reference materials Al, Ag, Au, Sn, and In. XRF measurements were performed with a Proto mScanXRF instrument; the composition distributions of elements are shown in Figure S3.

The calculation workflow integrated Pymatgen for composition handling and phase diagram construction, Matminer for feature generation, and Scikit-learn for machine learning (ML) model construction. The energy above the convex hull ($e_{above\ hull}$) for Fe-Ni-Mn-Cr alloys was predicted using a hybrid DFT-ML approach. The DFT dataset of formation energies from a previous high-throughput DFT study was used, with a few additions of intermetallics phases relevant to this work. The DFT dataset mainly comprises intermetallics structures and disordered phases represented by special quasi-random structures. Based on the DFT results, the most stable intermetallic phases on the convex hull can be identified and $e_{above\ hull}$ for disordered phases can be calculated for the compositions considered in the DFT dataset. The CALPHAD simulation was performed using the computation software, PanDat™ [37], for equilibrium phases at 500 - 700 °C, in which the composition space was mapped in 5 at% increments from 0-100 at% of each element [18]. To obtain results for compositions not considered in either DFT or CALPHAD datasets, ML models were trained on DFT-calculated formation energies of disordered phases and CALPHAD-calculated phase fractions, respectively. Compositional features were generated using Matminer, including stoichiometric ratios (e.g., atomic fractions), Magpie elemental properties (electronegativity, radii), valence orbital occupancies, and ionic characteristics, yielding a 145-dimensional feature vector. After feature correlation analysis with a threshold of 0.8 (Figures S14, S15), 21 features were selected for further training. ML models (Random Forest, XGBoost, and Gaussian Process) were trained separately for BCC and FCC lattices. Random Forest employed 500 trees with max_features=1/3, while XGBoost used 1000 trees (learning_rate=0.05, max_depth=6). Models were evaluated via 10-fold cross-validation (Table S3), and final predictions were made on a quaternary composition grid (Fe-Ni-Mn-Cr, 2 at% step, Figure S8c) with fixed Fe:Ni (1:1). Transition temperatures were estimated by $T_{trans} = \frac{e_{above\ hull}}{k_B S_{conf}}$, enabling high-throughput screening of metastable phases.

Acknowledgment

This research is part of the University of Toronto's Acceleration Consortium, which receives funding from the Canada First Research Excellence Fund (CFREF) and Natural Sciences and Engineering Research Council of Canada (NSERC). Part of the research described in this paper was performed at the Canadian Light Source (CLS), a national research facility of the University of Saskatchewan, which is supported by the Canada Foundation for Innovation (CFI), the Natural Sciences and Engineering Research Council (NSERC), the National Research Council (NRC), the Canadian Institutes of Health Research (CIHR), the Government of Saskatchewan, and the University of Saskatchewan

The authors thank Adam Leontowich (Canadian Light Source) for the experimental support.

Conflict interests

There are no conflicts of interest to declare.

Data Availability

The authors confirm that the data supporting the findings of this study are available within the article and its supplementary materials.

**Supplementary Information**

**High-throughput validation of phase formability and simulation accuracy of Cantor alloys**

Changjun Cheng[1,&], Daniel Persaud[1,&], Kangming Li[3], Michael J. Moorehead[4], Natalie Page[5], Christian Lavoie[6], Beatriz Diaz Moreno[7], Adrien Couet[8], Samuel E Lofland[5], Jason Hattrick-Simpers[1,2,9,10*]

1 Department of Materials Science and Engineering, University of Toronto, Toronto, ON, Canada

2 Acceleration Consortium, University of Toronto, Toronto, ON, Canada

3 Material Science and Engineering, King Abdullah University of Science and Technology, Thuwal, Kingdom of Saudi Arabia

4 Idaho National Laboratory, Idaho Falls, ID, United States

5 Department of Physics and Astronomy, Rowan University, Glassboro, NJ, United States

6 IBM T.J. Watson Research Center, Yorktown Heights, NY, United States

7 Canadian Light Source, Saskatoon, SK, Canada

8 University of Wisconsin-Madison, Madison, WI, United States

9 Schwartz Reisman Institute for Technology and Society, Toronto, ON, Canada

10 Vector Institute for Artificial Intelligence, Toronto, ON, Canada

& The authors contributed equally to this work

* Corresponding author: jason.hattrick.simpers@utoronto.ca

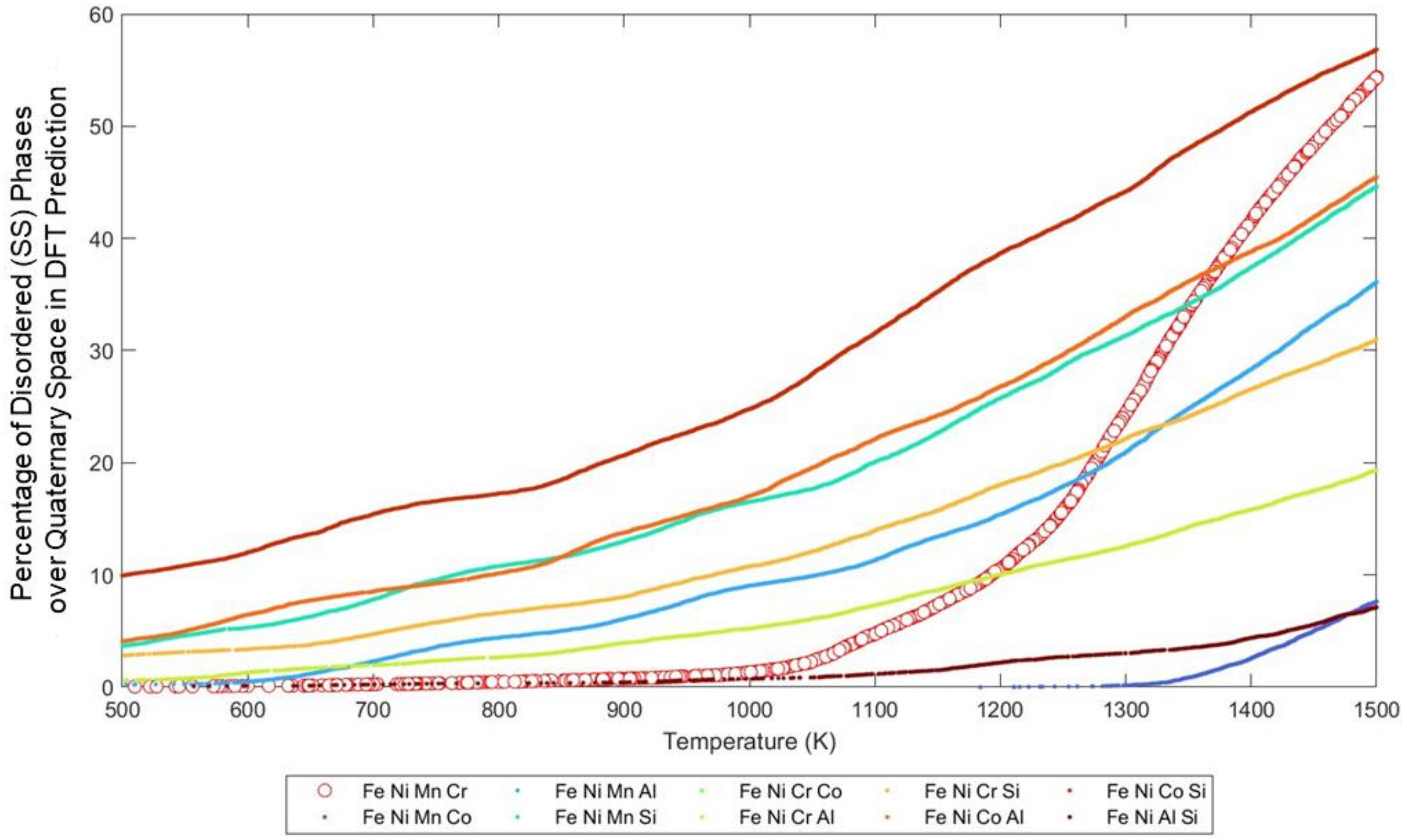

Figure S1. High-throughput screening on quaternary space as a function of SS-IM transformation temperature. Based on the Random Forest model trained on DFT input of energy above hull, the corresponding ordered intermetallic to disordered solid-solution (IM-SS) transformation temperatures were predicted over Fe-Ni-Mn-Cr-Co-Al-Si space, highlighting critical phase evolution mechanisms in FeNiMnCr system.

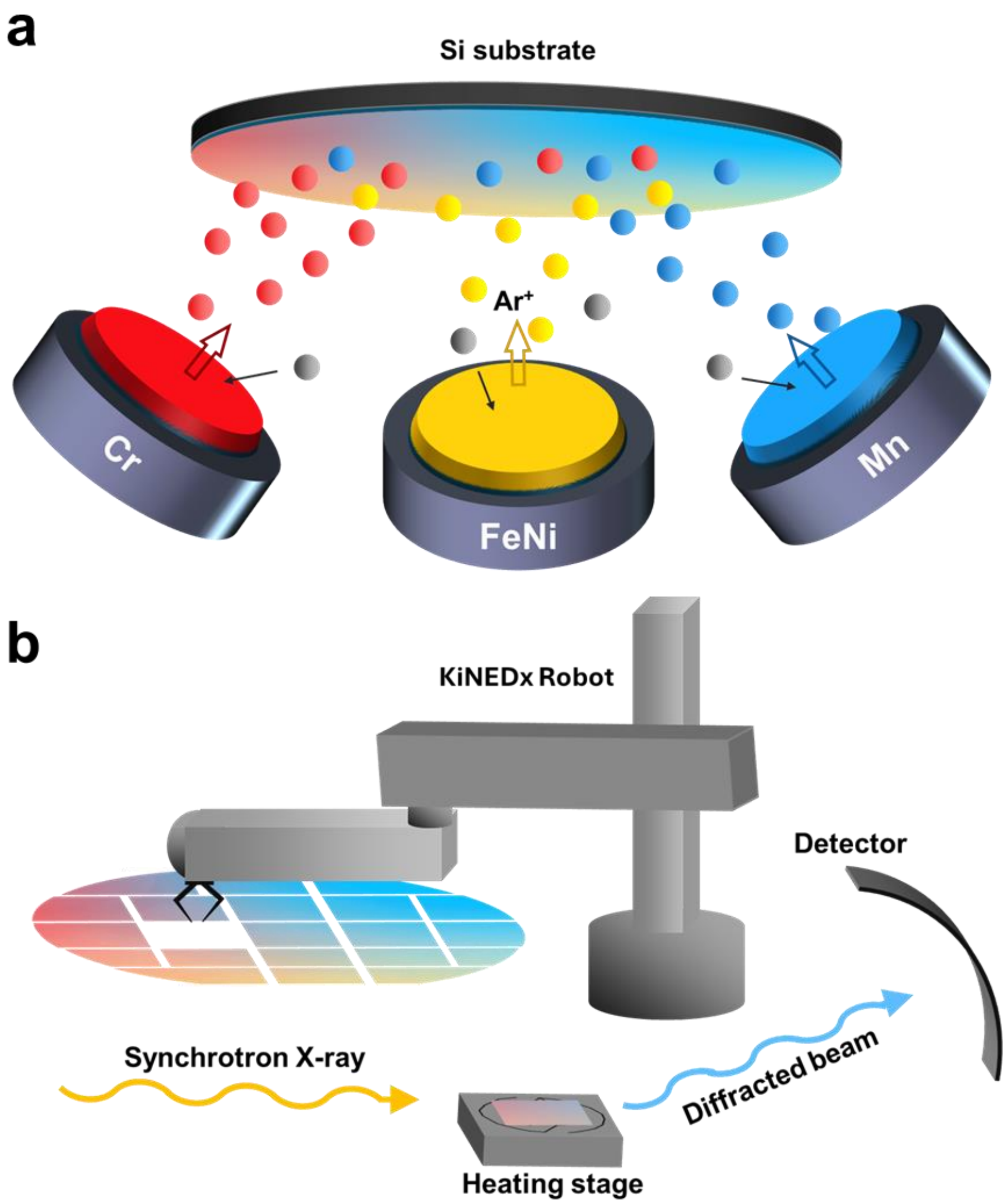

Figure S2. Schematics of (a) magnetron co-sputtering and (b) automated in-situ XRD diffraction.

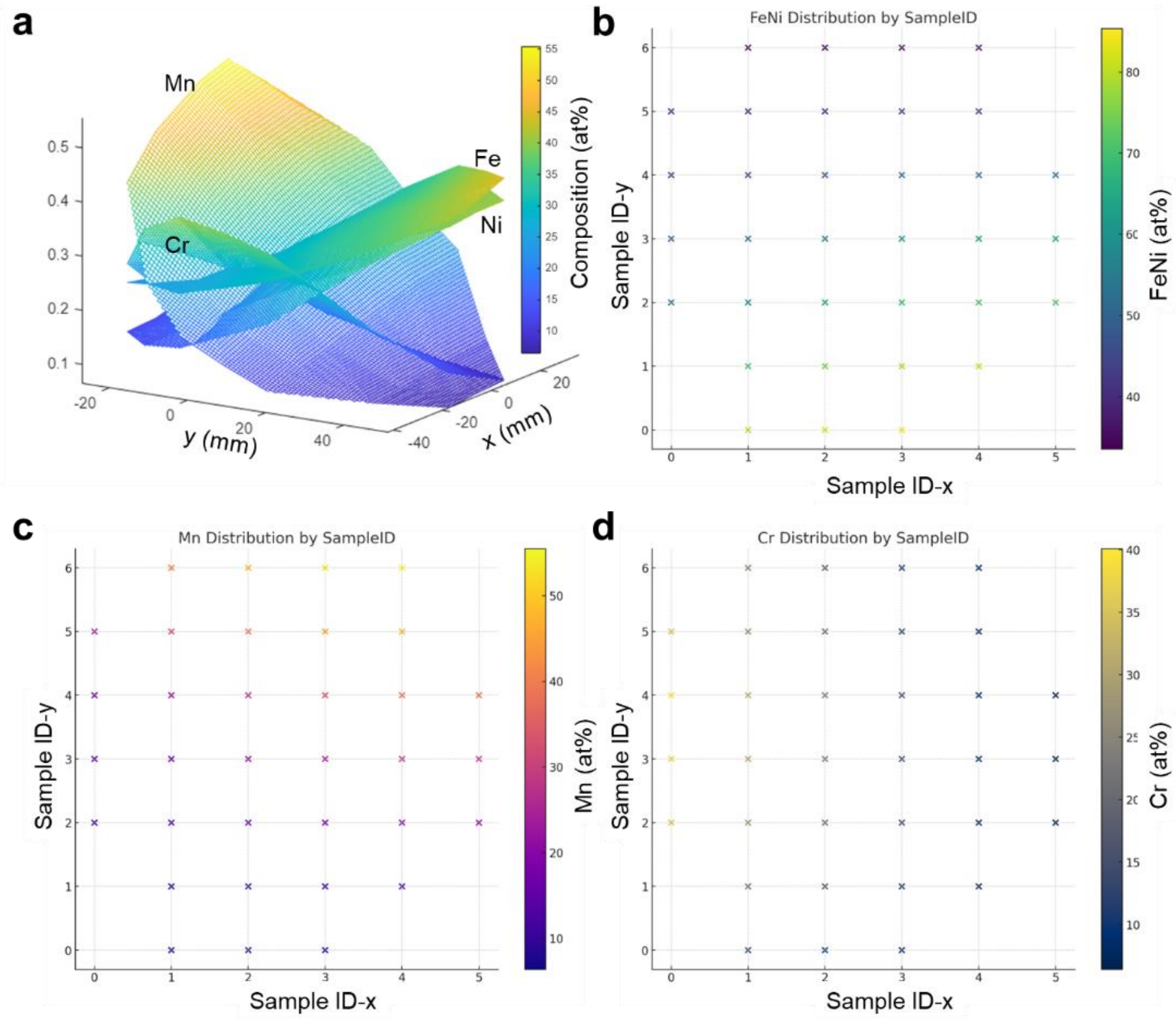

Figure S3. X-ray fluorescence results of element distribution on the wafer and corresponding compositions for the samples measured by in-situ XRD.

Figure S4 shows the detailed phase transformation from 600 °C to 1000 °C. Considering the onset of silicide appearance (Figure S4a), the enrichment of Mn leads to a significant decrease in formation temperature (< 600 °C). All samples experienced a step-by-step reaction to silicon, consequently forming stable transition metal silicide (TM-Si, $P2_13$), where Fe, Ni, Mn, Cr elements are randomly distributed in the lattice. A simple reaction process occurred in the FeNi-rich region (Figure S4b): the start of Si influx led to the formation of $FeNiSi_4$; then it disappeared, and the stable TM-Si existed at the highest temperature (1000 °C). Similar reaction steps occurred in the FeNi/Cr transition samples (Figure S4c); $FeNiSi_4$ became stable, and $CrSi_2$ also appeared at 1000 °C. As Cr concentration increased (Figure S4d), the FeNi-rich phases disappeared; besides the existence of stable $CrSi_2$ and TM-Si as final products, FCC phase, $Mn_4Si_7$, and $Cr_3Si$ progressively formed and disappeared as intermediate products; the easier reaction between Mn and Si resulted in the appearance of $Mn_4Si_7$ at lower temperatures (< 700 °C). In the Mn-rich region (Figure S4e), $Mn_4Si_7$ and TM-Si became the final products; besides the early Mn-Si reaction, dual BCC and intermetallic $FeMn_2$ phases also formed at medium temperatures. The Mn/Cr transition samples (Figure S4f) exhibited a combination of reactions in Mn-rich and Cr-rich samples. It should be noted that the FCC phase may also exist in Mn-rich and Cr-rich regions; however, the peaks overlapping with Mn-rich BCC and $Cr_3Si$ phases lead to the difficulty in clearly identifying the FCC phase. The silicide formation tendency can also be evaluated using the DFT method (Figure S4), demonstrating the earlier formation of $Cr_3Si$ and $Mn_4Si_7$ and the existence of final TM-Si products.

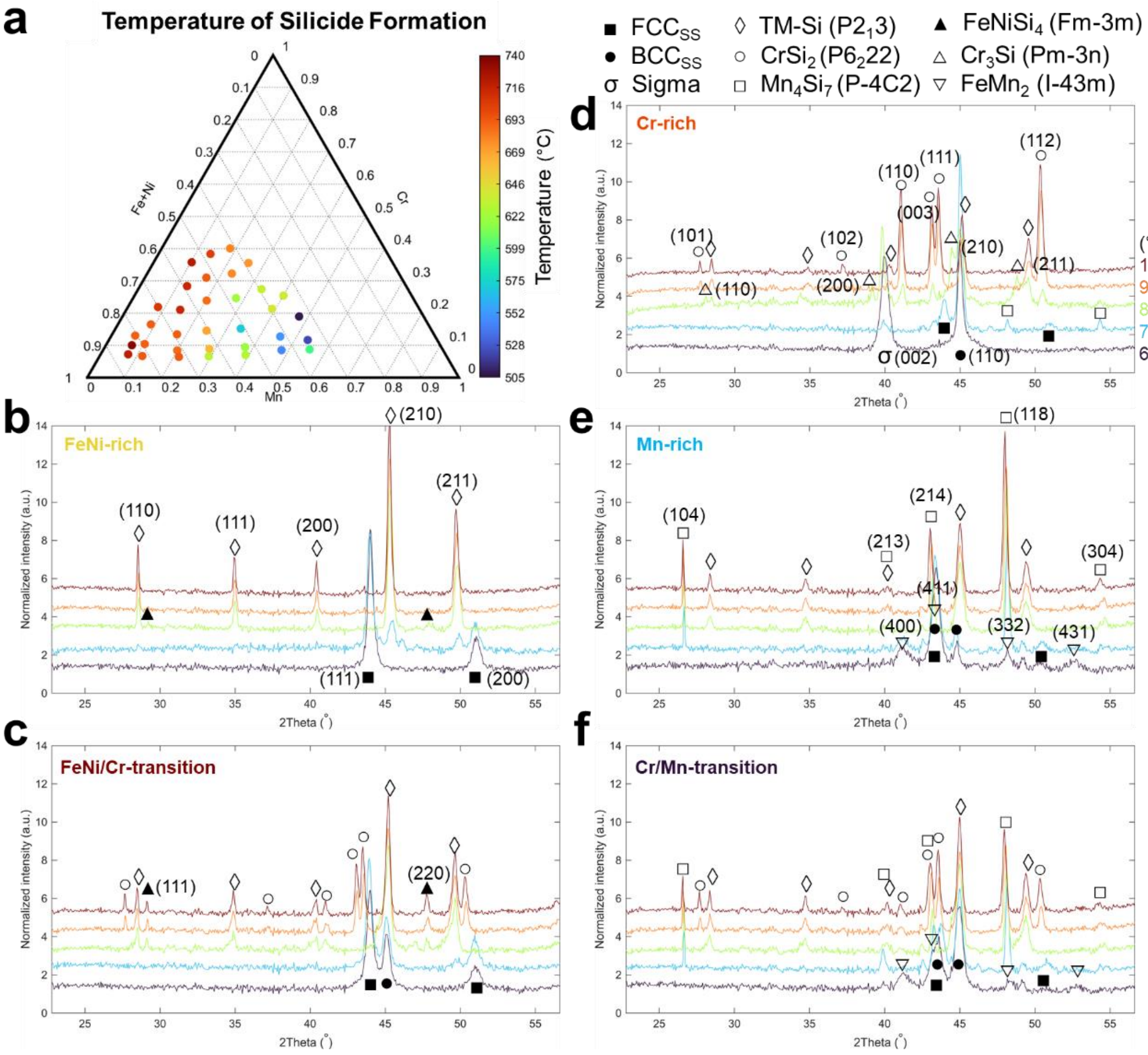

Figure S4. Silicidation reactions of Cantor library. (a) The temperature of silicide formation. (b) XRD curves of different samples in the temperature range between 600-1000 °C and corresponding product phases. Two theta values are presented in Cu-Kα wavelength.

Potential phases of silicide structures were identified through XRD. Their free energies were evaluated by considering the ground state energies at 0 K and the ideal configurational entropy: $G = H_{0\,k} - T * S_{conf}^{ideal}$. Under this assumption, the free energy change is attributed to the entropic contribution. For intermetallics (silicide), their configuration entropy is zero, so the free energy is constant with respect to temperature. In contrast, FeNiSi4, as well as the final TM-Si products, are partially disordered, with Fe and Ni (Fe, Ni, Mn, and Cr) randomly distributed in the same sublattice. Therefore, ideal configurational entropy on the corresponding sublattice is considered.

In the FeNi-rich sample, $FeNiSi_4$ is more likely to form first and then transform into TM-Si due to the considerable free energy decrease of TM-Si at higher temperatures. With more Cr added to the system, $CrSi_2$, with relatively high free energy, forms at higher temperatures than $FeNiSi_4$ and TM-Si in the FeNi/Cr-transition region. In the Cr-rich sample, metastable $Cr_3Si$ and $Mn_4Si_7$, with significantly low free energy, are the earlier products. The energy difference between the two intermetallics ($Cr_3Si$ and $Mn_4Si_7$) is only 0.006 eV/atom, which is close to DFT accuracy limit, especially when considering there could be temperature-induced effects on relative stability. With more Mn added to the system, the reaction follows the sequence of silicide with low to high free energy, i.e., $Mn_4Si_7$ to TM-Si to $CrSi_2$ in the Cr/Mn-transition region; a similar process occurs in the Mn-rich sample without the formation of $CrSi_2$.

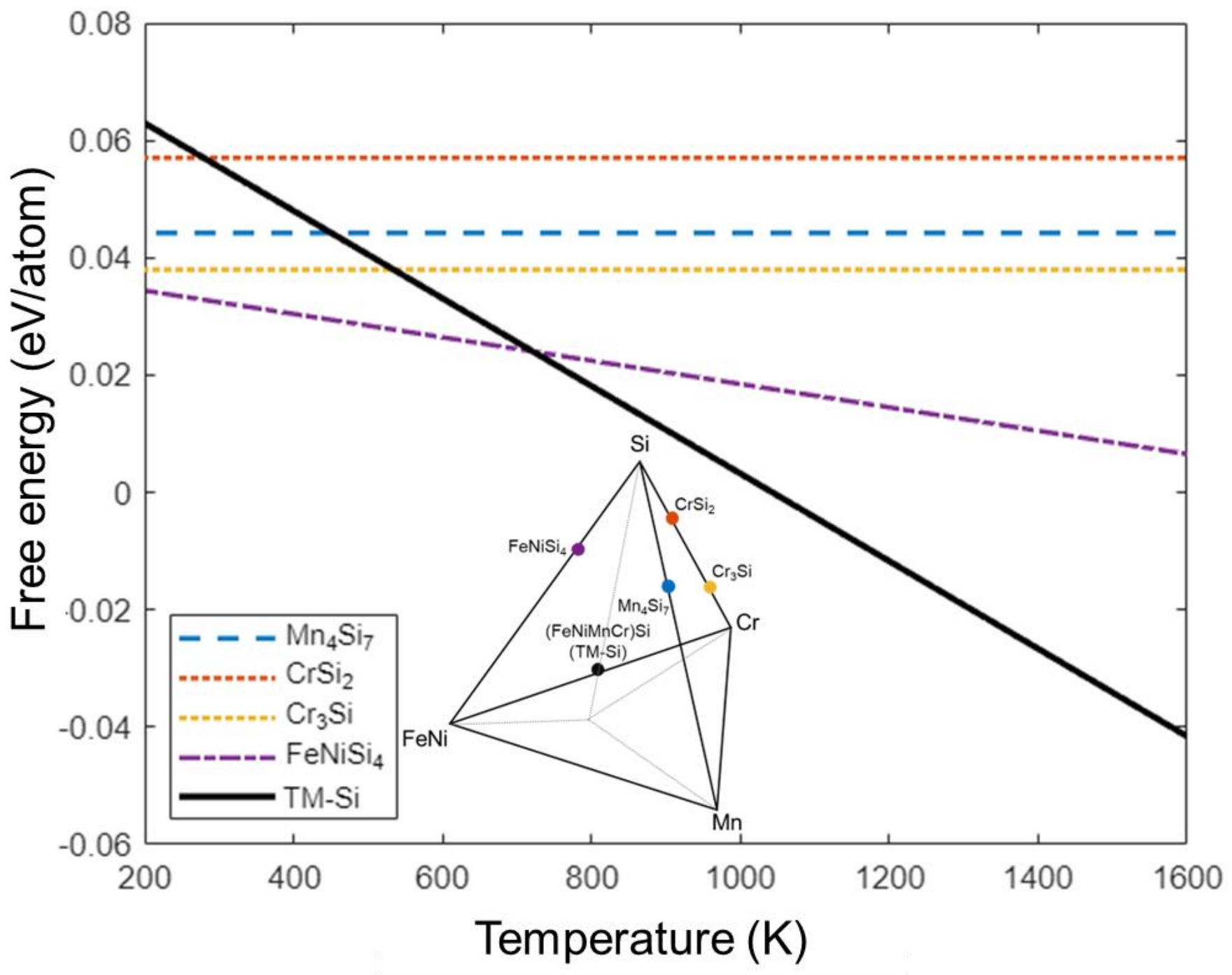

Figure S5. Comparison of formation tendency between metastable silicide and final product TM-Si, in which the free energy is calculated with respect to the convex hull energy

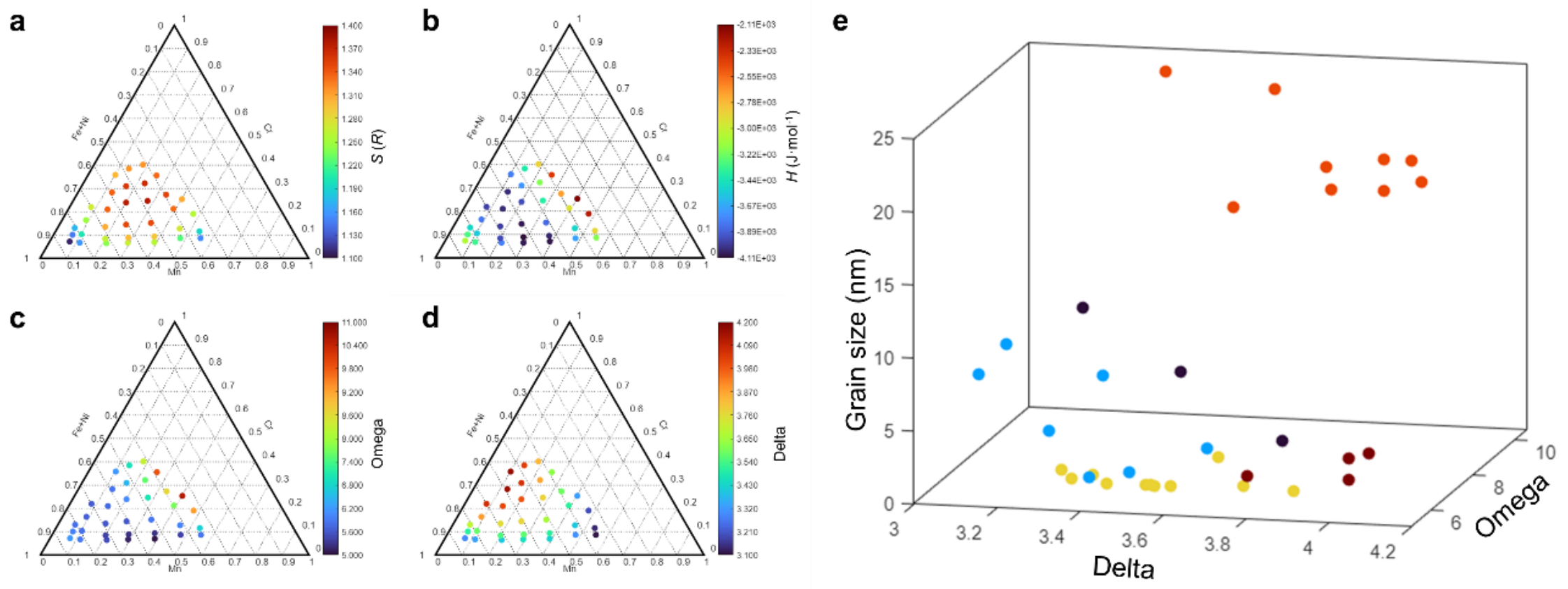

Figure S6. The values of (a) configuration entropy, (b) mixing enthalpy, (c) Omega, and (d) atomic radii mismatch. (e) The distribution of different-type samples in the Grain size-Omega-Delta plot.

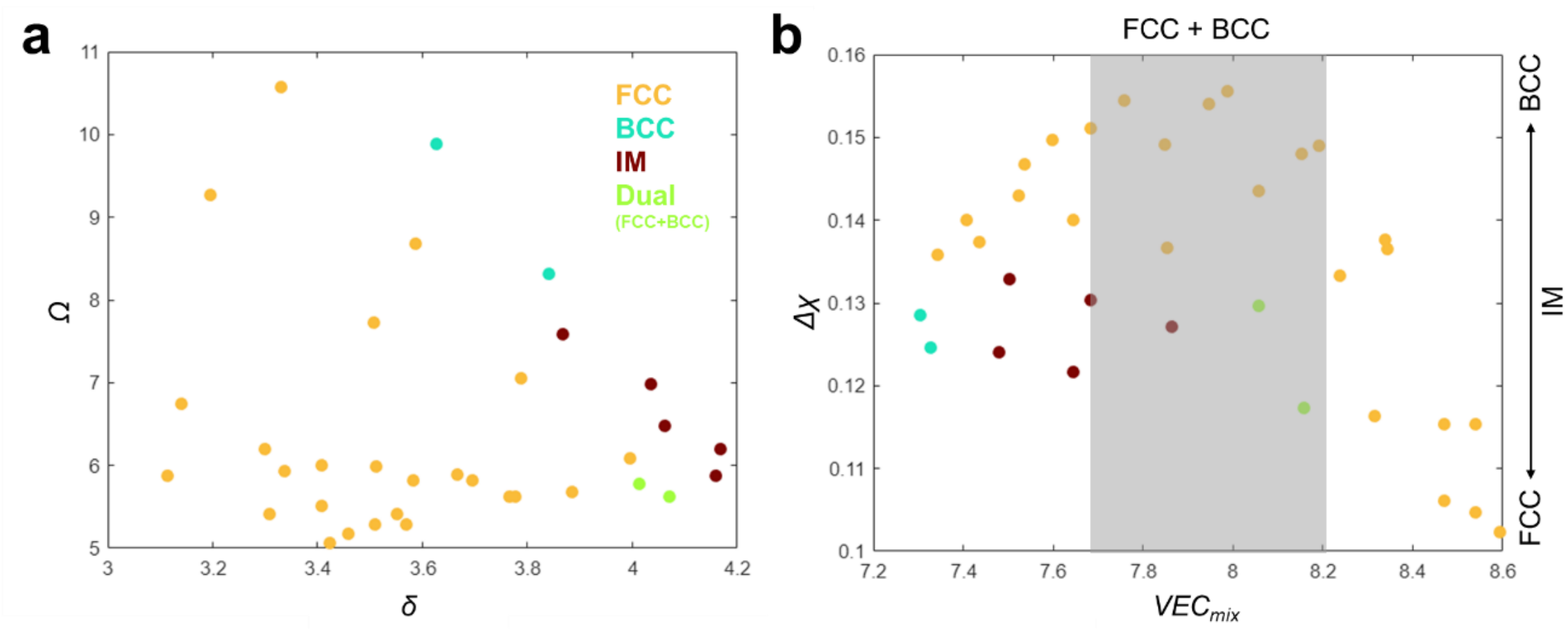

Figure S7. Based on the DFT-ML model, (a) the distribution of different phase types in the *Ω*-*δ* plot and (b) the distribution of phase types in *Δχ*-$VEC_{mix}$ plot.

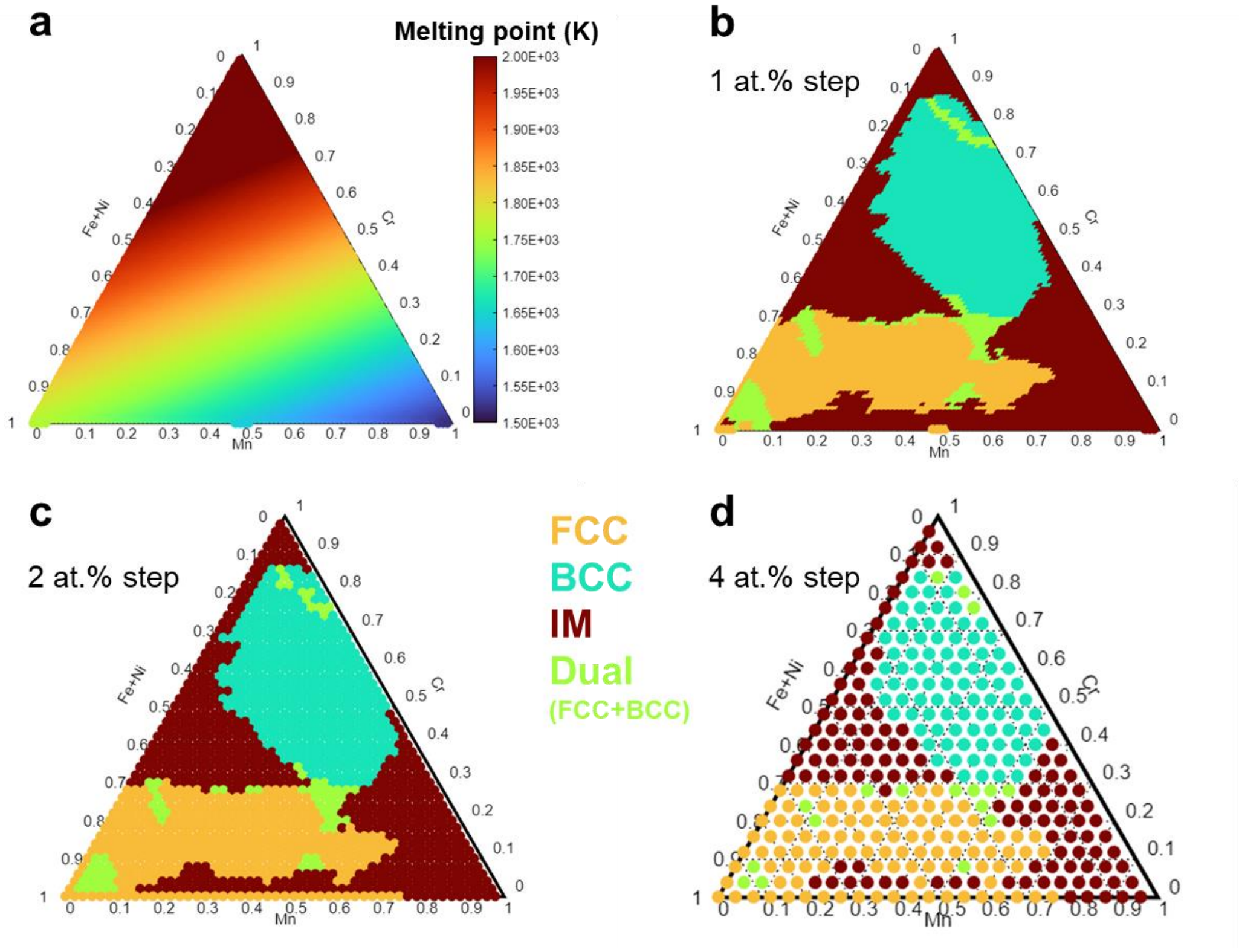

Figure S8. (a) Calculated melting points for FeNi-Mn-Cr HEAs based on the rule of mixtures. (b-c) Phase predictions using the RF model with different composition resolutions.

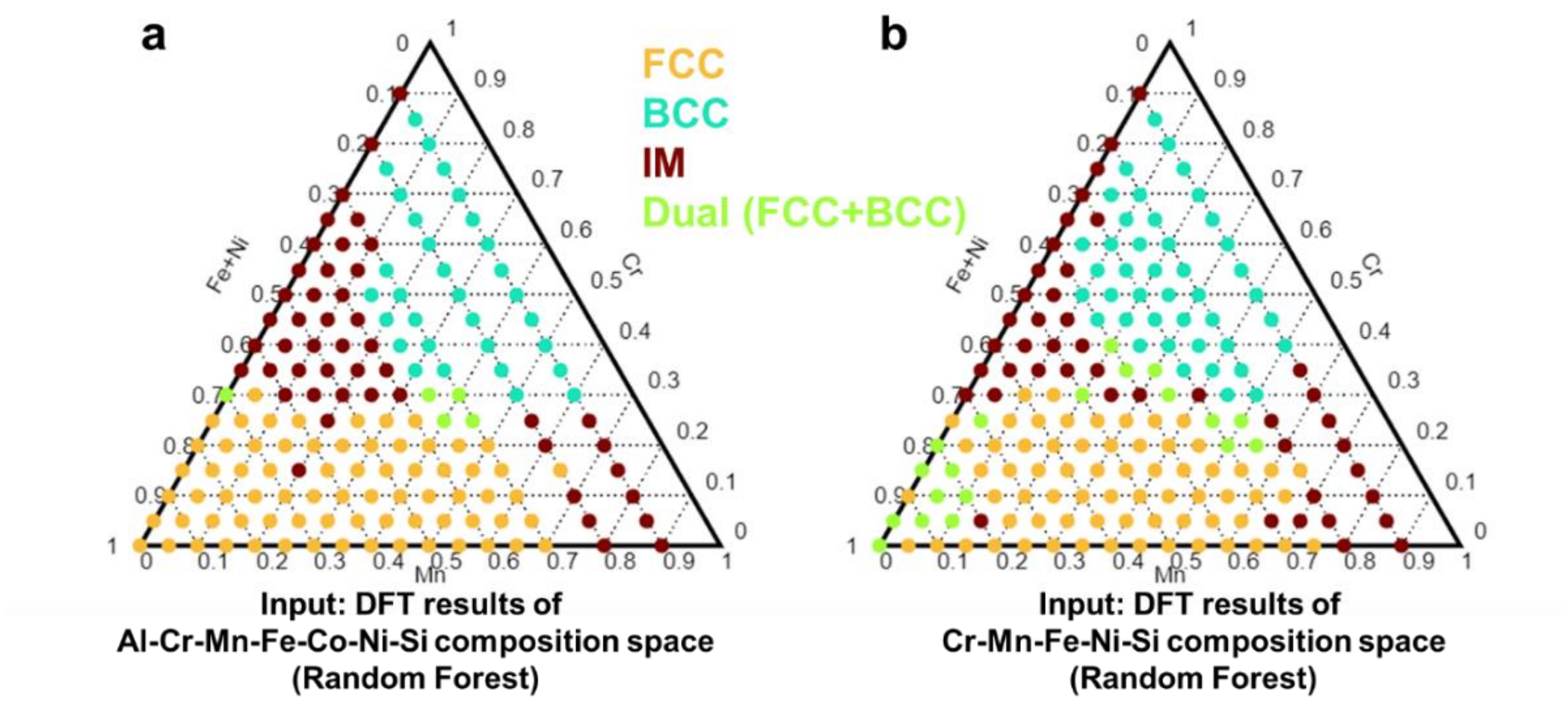

Figure S9. Comparison of phase predictions using the RF model with input of (a) 7-element and (b) 5-element composition spaces. The raw DFT calculation inputs largely affect the ML prediction. Despite the model presenting the FeNi/Cr-transition region with a dual phase of FCC+BCC structures, it mistakenly predicts the FeNi-rich region with also dual phase (Figure S8a); for the inputs with a larger element space, i.e., more Fe- and Ni-rich data provided in other alloy systems, the dual phase region significantly disappears (Figure S8b).

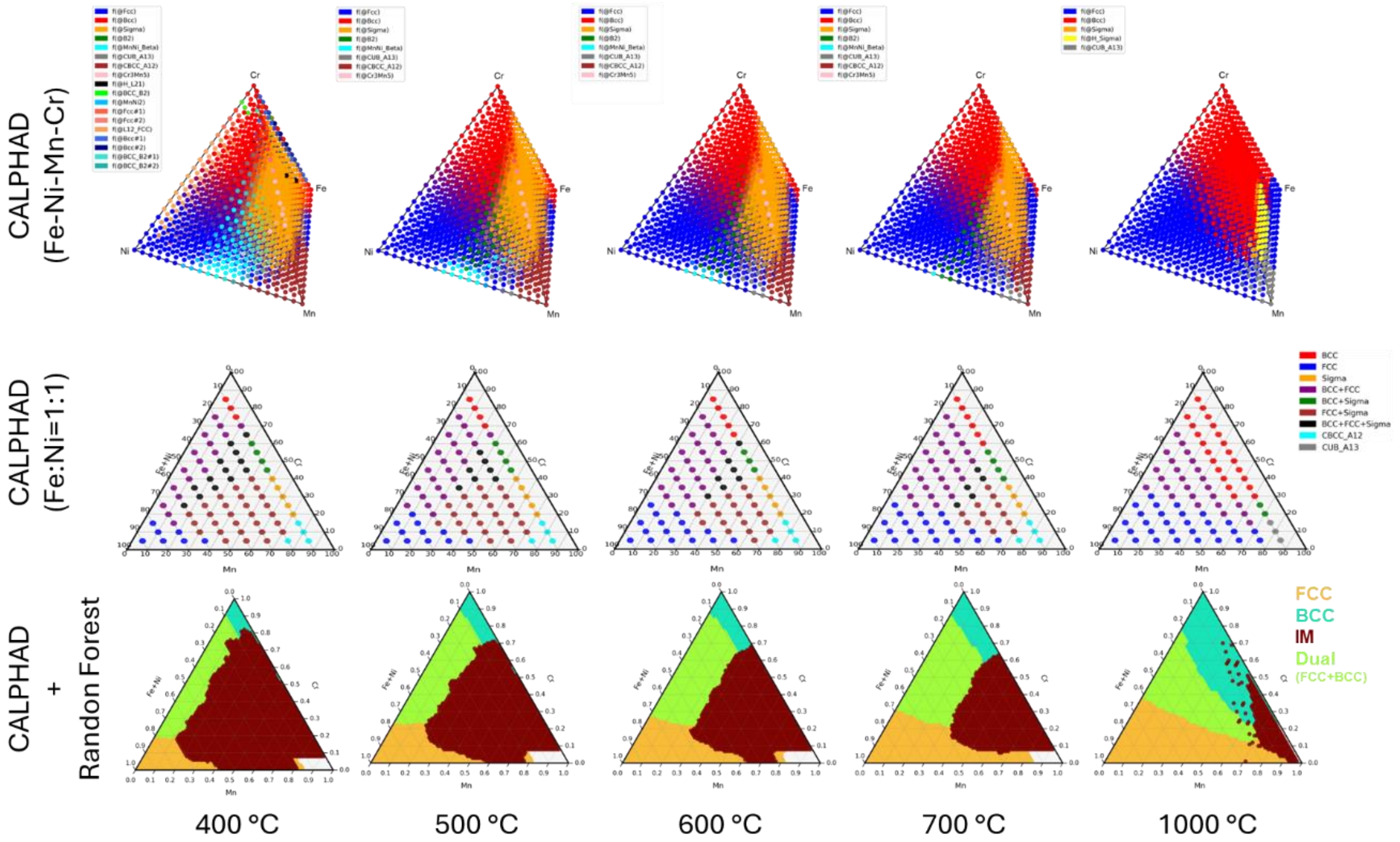

Figure S10. CALPHAD results at 400-700 °C and 1000 °C for (a) the whole Fe-Ni-Mn-Cr HEA space and (b) the ternary diagram with Fe:Ni = 1:1, as well as (c) the phase prediction based on the hybrid CALPHAD/RF model.

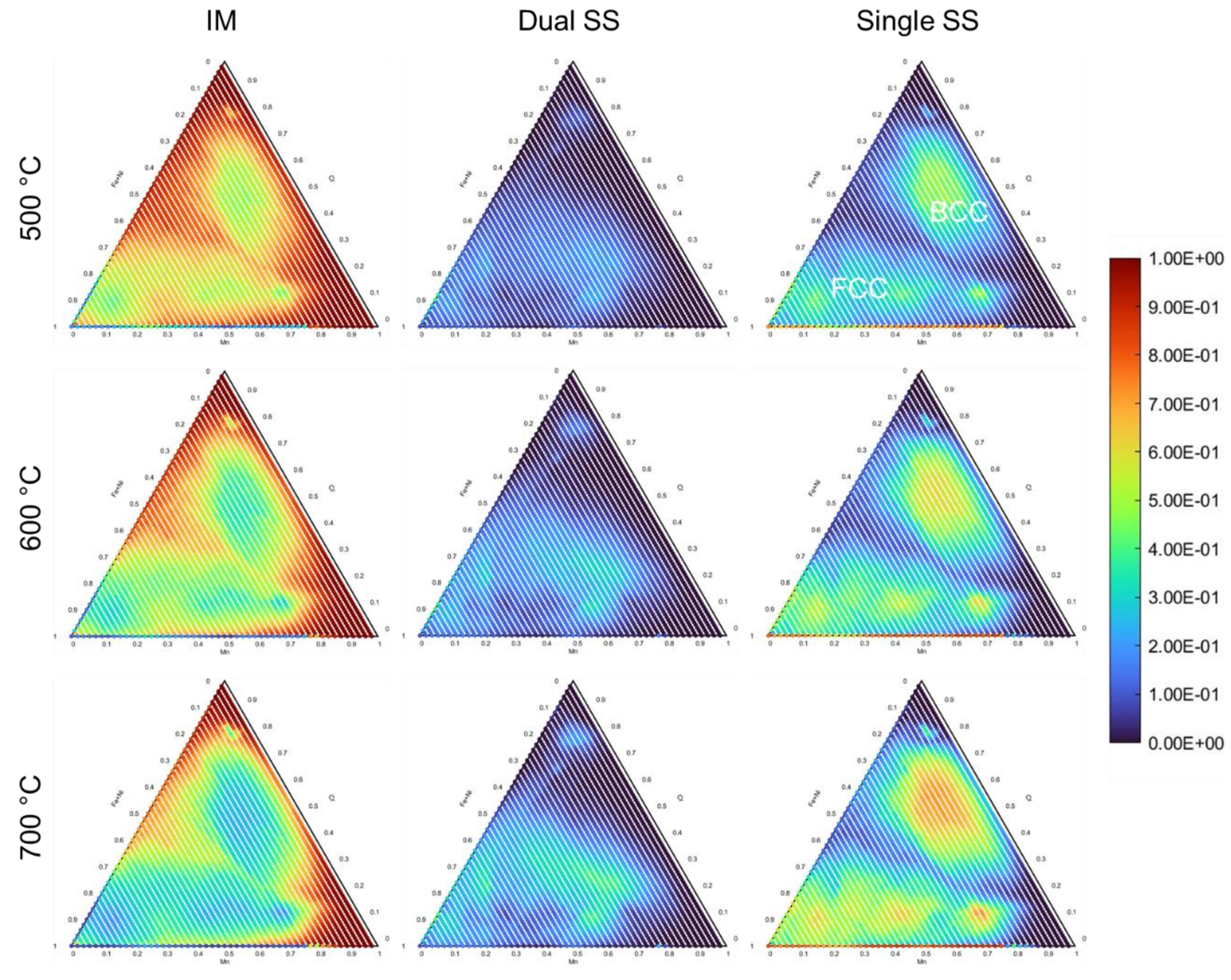

Figure S11. The formability of IM, dual, and single SS phases at 500-700 °C based on the modified sigmoid calculation on DFT-ML results

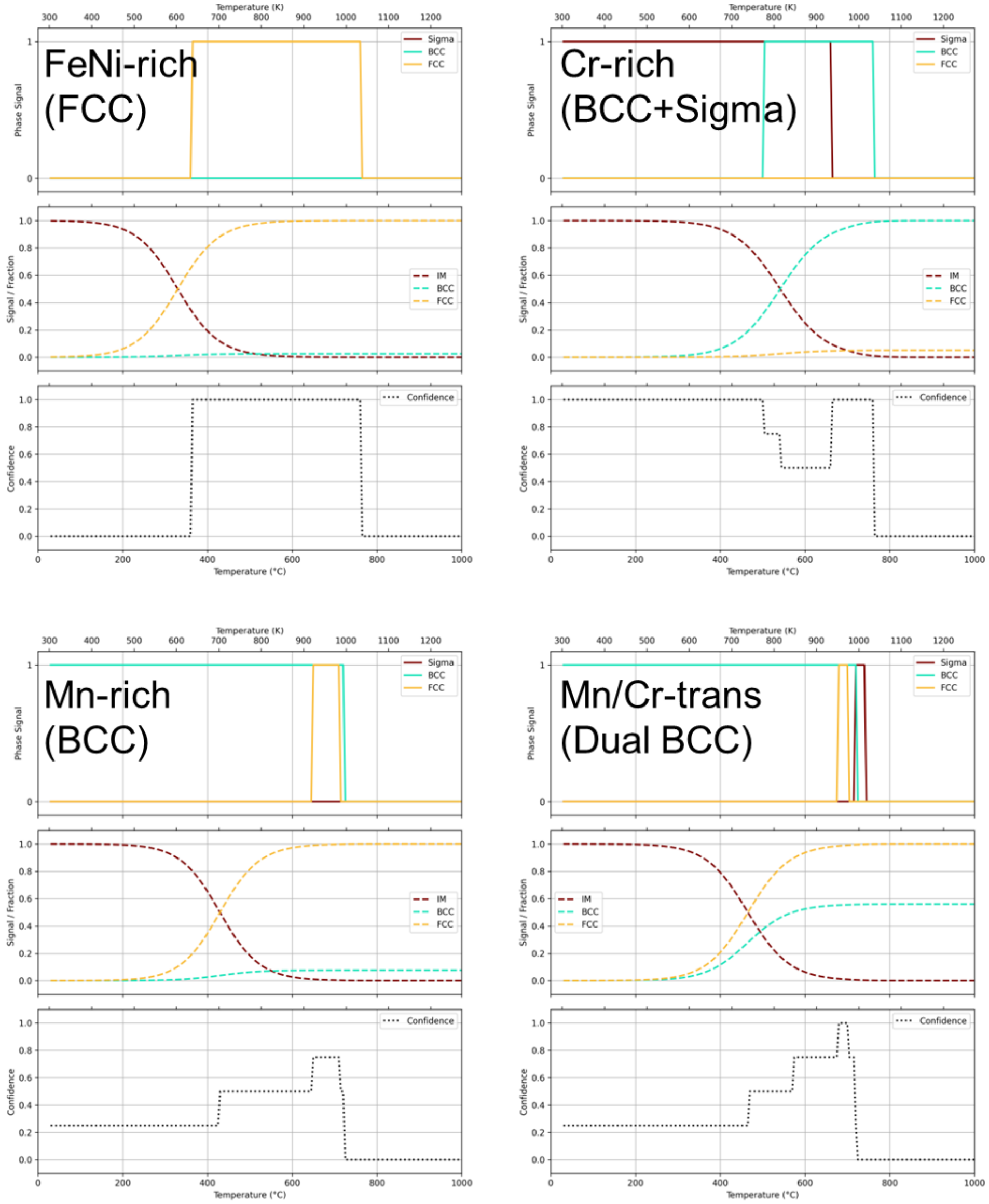

Figure S12. The comparison between XRD-indexed and DFT-ML predicted phases, as well as the corresponding confidence values as a function of temperature for the samples in Fe-rich, Cr-rich, Mn-rich, and Mn/Cr-transition regions.

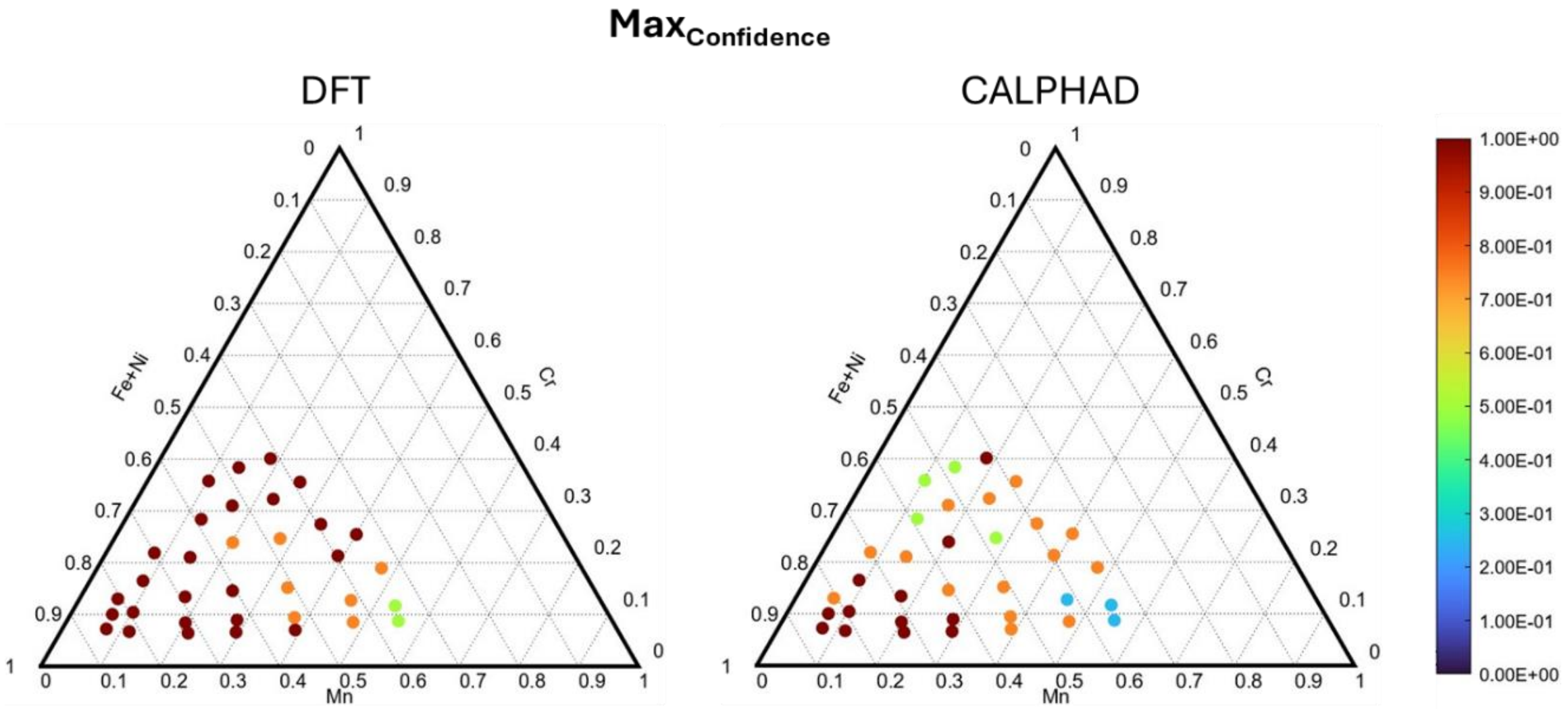

Figure S13. The maximum confidence values of DFT-ML and CALPHAD-ML models in the range of 500-700 °C. The overall high values of $Max_{Confidence}$ in the DFT-ML model suggest that the threshold parameters were well selected in the modified sigmoid function. The $Max_{Confidence}$ in the CALPHAD-ML model presents the prediction accuracy at 700 °C.

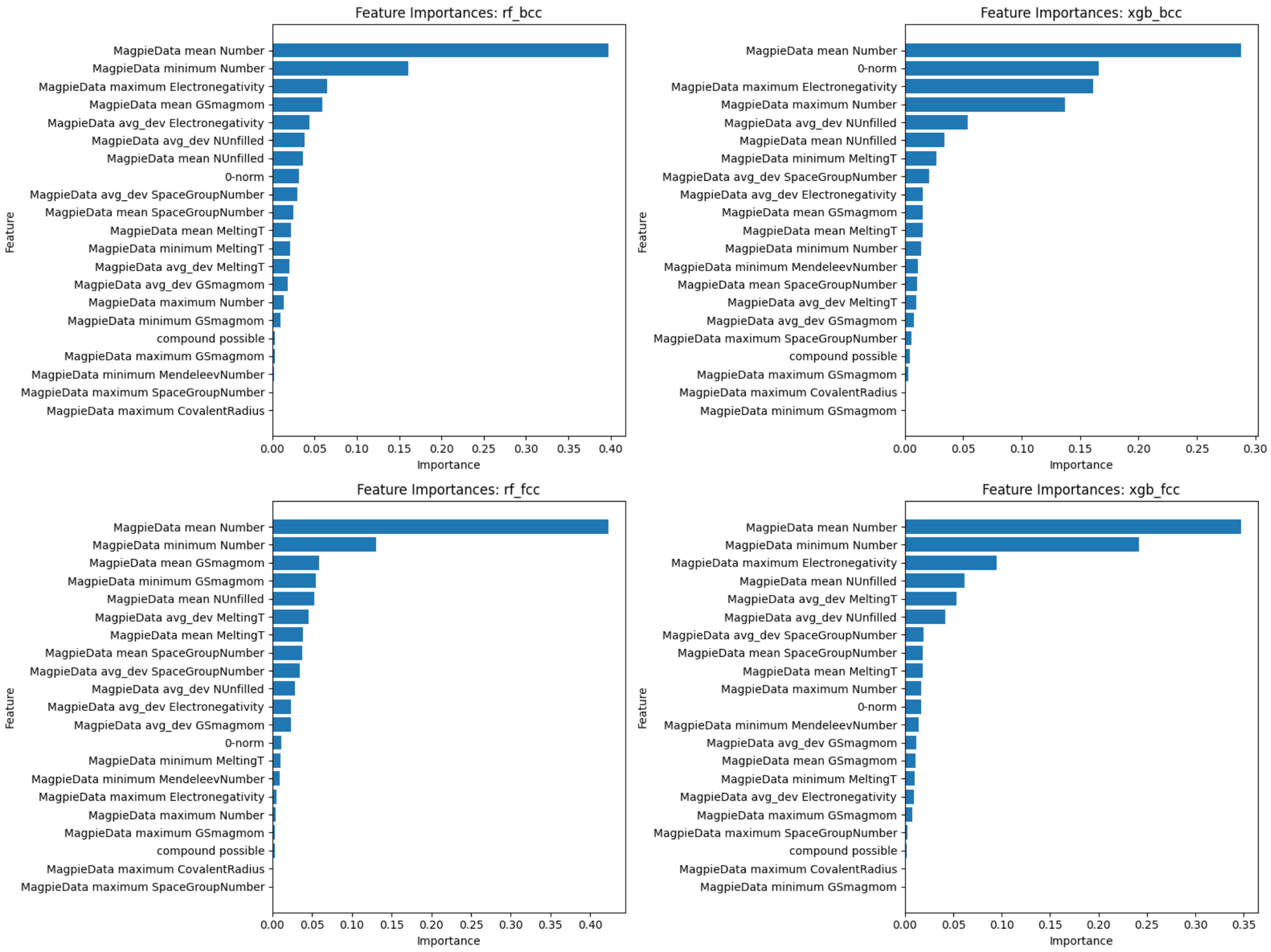

Figure S14. Feature importances of RF and XGBoost models for BCC and FCC predictions.

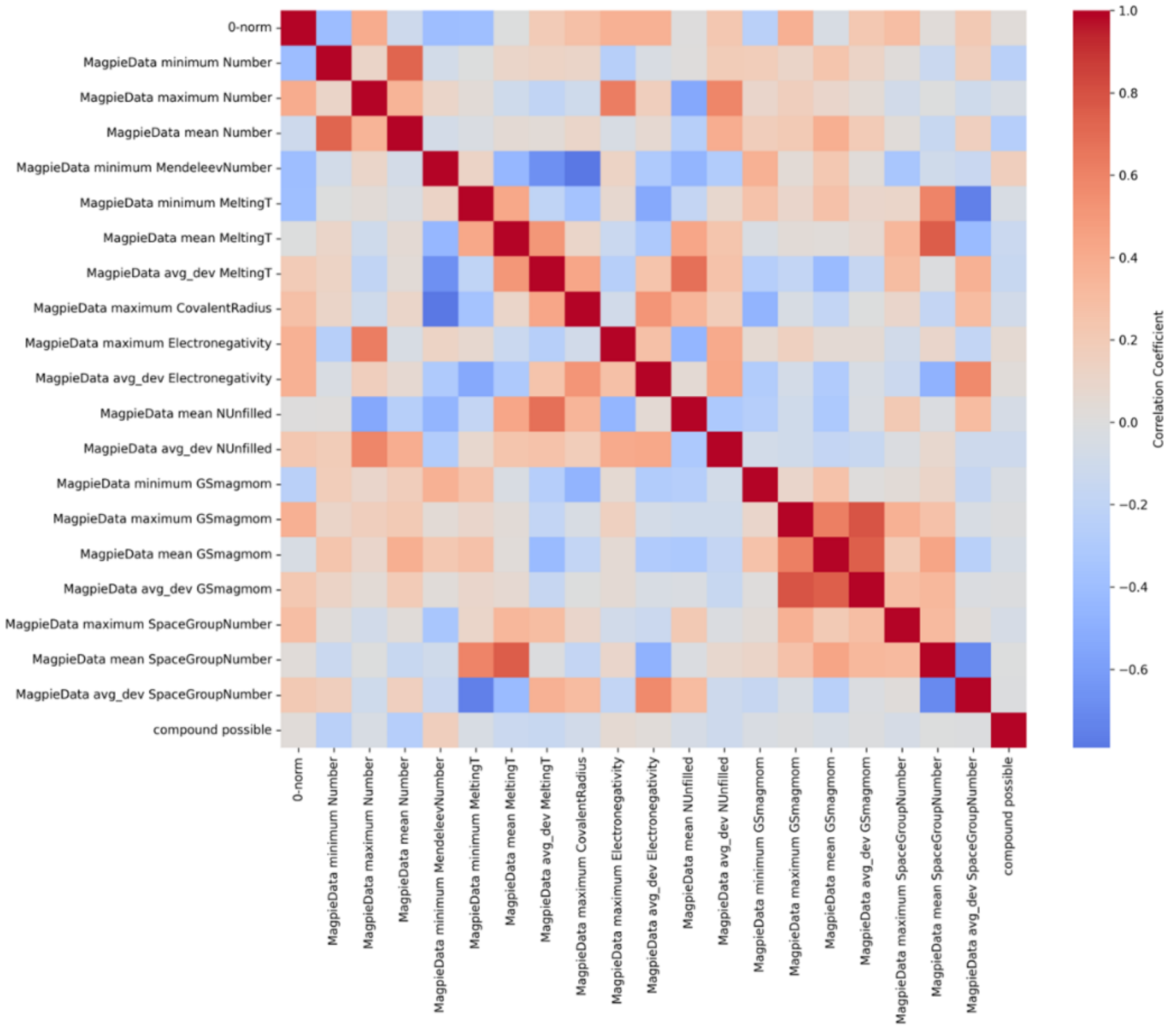

Figure S15. Correlation matrix of reduced magpie features.

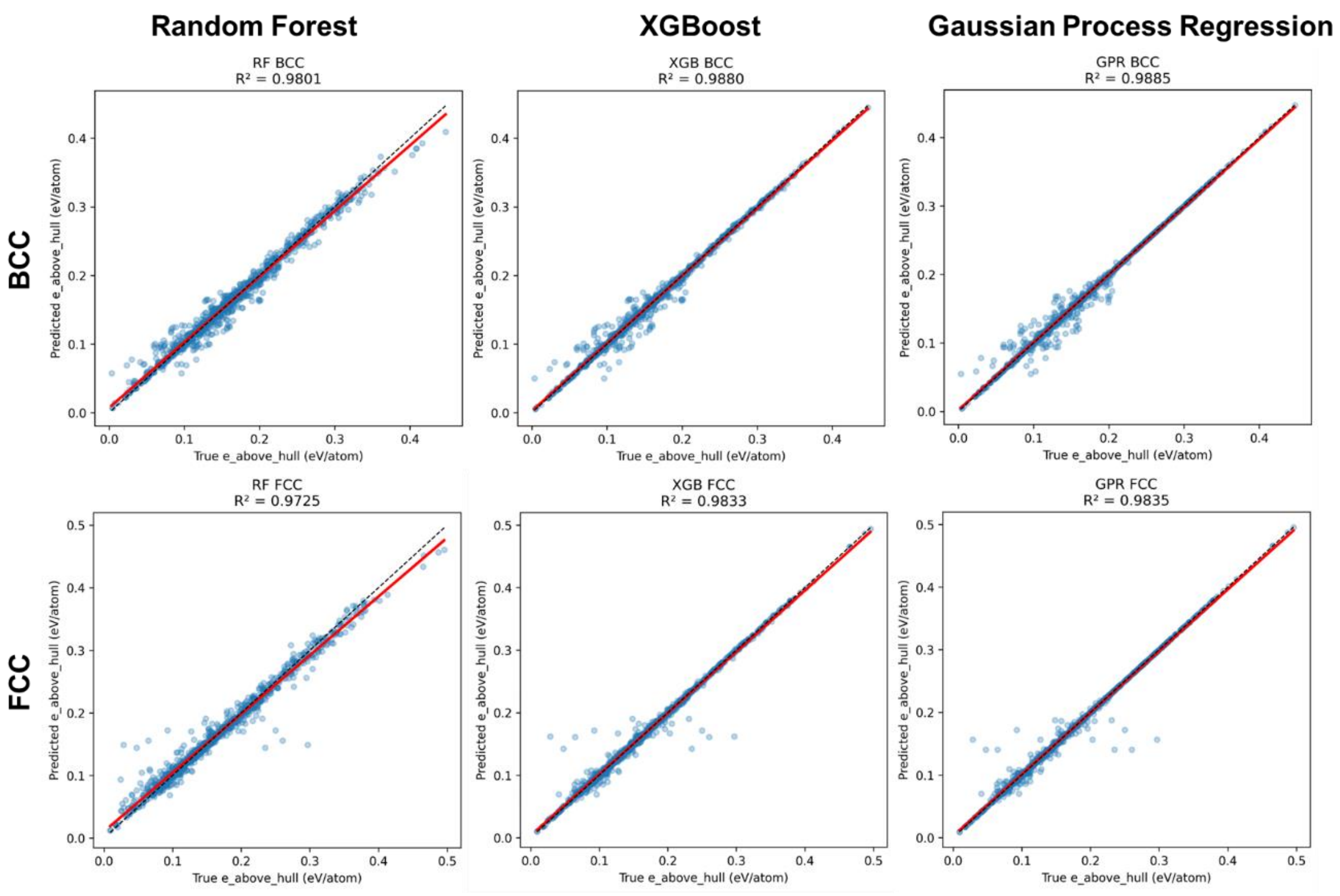

Figure S16. Comparison of ML model performance on BCC and FCC phase predictions using the whole DFT results as input with reduced magpie features.

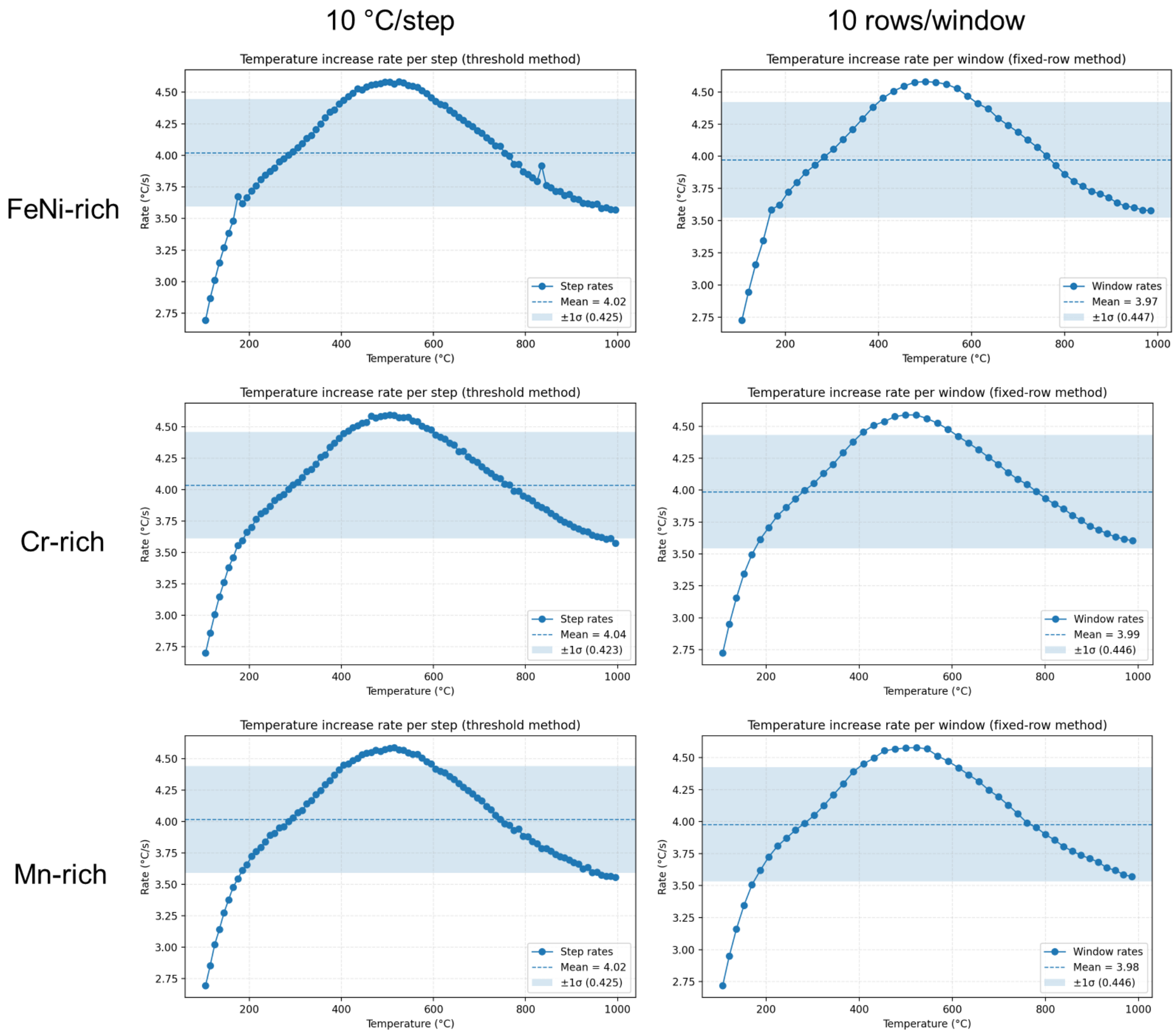

Figure S17. Temperature ramp rate variation within the range of 100-1000 °C. The collected transition points around 500 °C experienced a slight shift to higher values compared with the points at other temperatures.

The calculation of confidence level is based on the equation:

$$Confidence = \frac{1}{n} * \sum_{i}^{\frac{BCC}{FCC}, \frac{SS}{IM}} A_i \;\; (n = 2, A_i = 0, 0.5, or\ 1). ,$$

where $A_i$ is the agreement of phase formation prediction between the observed experiment and calculated results.

Table S1. Deposition parameters of magnetron sputtering for FeNiMnCr HEA library.

| | Avg Power (W) | Avg Voltage (V) | Avg Current (A) | Power Source |
|---|---|---|---|---|
| Fe/Ni | 30 | 341 | 0.073 | Advanced Energy MDX-1K |
| Mn | 5 | 282 | 0.035 | Advanced Energy MDX-1K |
| Cr | 11 | 284 | 0.041 | AJA DCXS-750-4 |

Table S2. Composition of samples in Cr-rich region as well as corresponding FeNiMn/Cr ratios. Some of data from other regions are also listed. The FeNi/Cr ratios cannot perfectly distinguish between Cr-rich and Cr/Mn-transition regions.

| **Composition (at %)** | | | FeNi/Cr | **FeNiMn/Cr** | Type |
|---|---|---|---|---|---|
| Fe+Ni | Mn | Cr | | | |
| 41.51 | 18.4 | 40.09 | 1.04 | **1.49** | Cr-rich (with σ phase) |
| 47.67 | 13.97 | 38.36 | 1.24 | **1.61** | |
| 54.01 | 10.22 | 35.77 | 1.51 | **1.80** | |
| 38.83 | 25.6 | 35.56 | 1.09 | **1.81** | |
| 44.95 | 22.77 | 32.28 | 1.39 | **2.10** | |
| 52.45 | 16.56 | 31 | 1.69 | **2.23** | |
| 58.94 | 12.68 | 28.38 | 2.08 | **2.52** | |
| 39.41 | 33.16 | 27.42 | 1.44 | **2.65** | |
| 34.45 | 40.08 | 25.46 | 1.35 | **2.93** | |
| 47.62 | 27.72 | 24.66 | 1.93 | 3.06 | Cr/Mn-transition |
| 55.93 | 20.18 | 23.9 | 2.34 | 3.18 | FeNi/Cr-transition |
| 70.02 | 8.09 | 21.89 | 3.20 | 3.57 | FeNi/Cr-transition |
| 39.61 | 39.06 | 21.33 | 1.86 | 3.69 | Cr/Mn-transition |
| 64.47 | 14.49 | 21.04 | 3.06 | 3.75 | FeNi/Cr-transition |
| 33.51 | 47.54 | 18.96 | 1.77 | 4.27 | Cr/Mn-transition |
| 74.61 | 8.88 | 16.5 | 4.52 | 5.06 | Fe-rich |

Table S3. Results of 10-fold cross-validation on ML models with 8 to 2 sample split. The performances of full magpie features and reduced magpie features are listed below.

| | Model | | CV_MAE_mean | CV_MAE_std | CV_R2_mean | CV_R2_std |
|---|---|---|---|---|---|---|
| DFT | Full magpie features | rf_bcc | 0.015232 | 0.001222 | 0.918039 | 0.013899 |
| | | xgb_bcc | 0.015645 | 0.00157 | 0.905876 | 0.020559 |
| | | gpr_bcc | 0.123 | 0.010739 | -3.55081 | 0.370172 |
| | | rf_fcc | 0.015298 | 0.002499 | 0.892795 | 0.040077 |
| | | xgb_fcc | 0.015803 | 0.003227 | 0.868284 | 0.063219 |
| | | gpr_fcc | 0.122067 | 0.009268 | -3.22886 | 0.504054 |
| | Reduced magpie features | rf_bcc | 0.01622 | 0.001186 | 0.907683 | 0.017522 |
| | | xgb_bcc | 0.016111 | 0.001659 | 0.90176 | 0.024405 |
| | | gpr_bcc | 0.123 | 0.010739 | -3.55081 | 0.370172 |
| | | rf_fcc | 0.017431 | 0.002429 | 0.872534 | 0.042845 |
| | | xgb_fcc | 0.016849 | 0.003035 | 0.8548 | 0.065498 |
| | | gpr_fcc | 0.122067 | 0.009268 | -3.22886 | 0.504054 |
| CALPHAD | Full magpie features | rf_bcc | 0.0139 | 0.0021 | 0.9804 | 0.0065 |
| | | xgb_bcc | 0.0123 | 0.0027 | 0.9851 | 0.0073 |
| | | gpr_bcc | 0.0248 | 0.0048 | 0.9205 | 0.0534 |
| | | rf_fcc | 0.0241 | 0.0040 | 0.9645 | 0.0109 |
| | | xgb_fcc | 0.0236 | 0.0044 | 0.9674 | 0.0135 |
| | | gpr_fcc | 0.0433 | 0.0068 | 0.8965 | 0.0446 |
| | | rf_sigma | 0.0185 | 0.0028 | 0.9773 | 0.0076 |
| | | xgb_sigma | 0.0172 | 0.0029 | 0.9798 | 0.0070 |
| | | gpr_sigma | 0.0214 | 0.0045 | 0.9569 | 0.0355 |
| | Reduced magpie features | rf_bcc | 0.018638 | 0.002745 | 0.975 | 0.006414 |
| | | xgb_bcc | 0.015056 | 0.001858 | 0.983173 | 0.003529 |
| | | gpr_bcc | 0.11371 | 0.055091 | 0.051356 | 0.574248 |
| | | rf_fcc | 0.027577 | 0.003919 | 0.964841 | 0.009086 |
| | | xgb_fcc | 0.025408 | 0.004946 | 0.968961 | 0.013374 |
| | | gpr_fcc | 0.045815 | 0.009525 | 0.889413 | 0.05011 |
| | | rf_sigma | 0.025713 | 0.002116 | 0.966544 | 0.006279 |
| | | xgb_sigma | 0.023882 | 0.002714 | 0.966568 | 0.010101 |
| | | gpr_sigma | 0.145656 | 0.012315 | -0.27334 | 0.02736 |